\documentclass[twocolumn,showpacs,prb]{revtex4}
\usepackage{amsmath,amssymb}

\usepackage{verbatim}

\usepackage{amssymb}
\usepackage{graphicx}
\usepackage{epstopdf}
\usepackage{amsmath}
\usepackage{amsfonts}

\usepackage{dcolumn}
\usepackage{bm}
\usepackage{hyperref}

\begin{document}

\title{Fermionic theory for quantum antiferromagnets with spin $S>1/2$}

\author{Zheng-Xin Liu$^{1}$}\email{liuzx@ust.hk}
\author{Yi Zhou$^{2}$}\email{yizhou@zju.edu.cn}
\author{Tai-Kai Ng$^{1}$}
\email{phtai@ust.hk}

\affiliation{1,Department of Physics,Hong Kong University of
Science
and Technology,Clear Water Bay Road, Kowloon, Hong Kong\\
2. Department of Physics, Zhejiang University, Hangzhou 310027, P.
R. China }

\begin{abstract}
The fermion representation for $S=1/2$ spins is generalized to
spins with arbitrary magnitudes. The symmetry properties of the
representation is analyzed where we find that the particle-hole
symmetry in the spinon Hilbert space of $S=$1/2 fermion
representation is absent for $S>1/2$. As a result, different path
integral representations and mean field theories can be formulated
for spin models. In particular, we construct a Lagrangian with
restored particle-hole symmetry, and apply the corresponding mean
field theory to one dimensional (1D) $S=1$ and $S=3/2$
antiferromagnetic Heisenberg models, with results that agree with
Haldane's conjecture. For a $S=1$ open chain, we show that
Majorana fermion edge states exist in our mean field theory. The
generalization to spins with arbitrary magnitude $S$ is discussed.
Our approach can be applied to higher dimensional spin systems. As
an example, we study the geometrically frustrated $S=1$ AFM on
triangular lattice. Two spin liquids with different pairing
symmetries are discussed: the gapped $p_x+ip_y$-wave spin liquid
and the gapless $f$-wave spin liquid. We compare our mean field
result with the experiment on NiGa$_2$S$_4$, which remains
disordered at low temperature and was proposed to be in a spin
liquid state. Our fermionic mean field theory provide a framework
to study $S>1/2$ spin liquids with fermionic spinon excitations.

\end{abstract}

\pacs{75.10.Kt, 75.10.Jm, 71.10.Hf}

\maketitle

\section{Introduction}
Quantum magnetism is one of the oldest central problems in
condensed matter and many body physics. Few exact results are
known except at one dimension or when long range spin order is
established at low temperature. In the later case, Landau's
symmetry breaking paradigm is applicable, and the long range spin
order determines the low energy physics of the system. The low
energy spin excitations are spin waves or magnons.\cite{spinwave}
Systematic corrections to spin wave theory can be obtained through
semiclassical approach ($1/S$ expansion\cite{1/S}) and/or
nonlinear $\sigma$ model.\cite{Haldane} However, it was suggested
that for some anti-ferromagnetic materials, strong quantum
fluctuations due to small spin magnitude and low dimensionality
combined with geometric frustration may lead to quantum coherent,
spin disordered ground states.\cite{AndersonSL, Leescience} In
such cases, a quantum fluid approach\cite{MF} is probably a better
starting point to describe the low energy physics instead of usual
spin wave or semiclassical approach.

There exist at present two common quantum fluid approaches based on
different ``particle" representations of spins. In the ``bosonic"
approach\cite{MF,Ma} spins are represented by spin up and down
Schwinger bosons with the constraint that total boson number at each
site is $2S$. A spin-disordered state is obtained in a mean field
theory as long as Bose condensation does not occur. The low lying
excitations coming from this theory are bosonic. The other
possibility is to represent spins by fermions. Up to now, the
fermionic approach is mainly restricted to $S=1/2$ spin
systems\cite{MF, AZHA-1/2}. Because of the Pauli exclusion
principle, we cannot put more than two spin up or down fermions on a
site to form $S>1/2$ objects as in bosonic approach. One way out is
to introduce more species of fermions. It was proposed in
Ref.~\onlinecite{JPA}, that $(2S+1)$-species of boson/fermions can
be used to construct the spin-$S$ swapping operators. Later, it is
shown that the fermionic representation can be used to describe a
$S=1$ valence bond solid state.\cite{PRB}

In this paper, we systematically study the fermionic representation
for arbitrary spins. To construct a spin operator with spin
magnitude $S$, we follow Ref.~\onlinecite{JPA} and introduce
different fermion operators for different $S_z$ states. We find that
for half-odd-integer spins, the spin operator thus constructed is
invariant under a $SU(2)$ transformations,\cite{AZHA-1/2} i.e. the
symmetry group is $SU(2)$ group whereas for integer spins the
symmetry group is $U(1)\bar\otimes Z_2$. Fermionic path integral
formulations for spin systems with arbitrary spin-$S$ can then be
formulated and the corresponding mean field theories can be studied.
We shall see that one of the mean field theories applied to 1D
antiferromagnetic Heisenberg model produces mean field excitation
spectrums which agrees with Haldane's conjecture.\cite{Haldane}
Moreover, there exist zero energy Majorana edge modes for open
integer spin chains. We also apply our approach to two-dimensional
(2D) antiferromagnetic Heisenberg models on triangular lattices
where we obtain two translational and rotational invariant spin
liquid solutions, one (with $p_x+ip_y$-wave paring symmetry) is
gapped and the other one (with $f$-wave pairing symmetry) has nodes
at the spinon Fermi level. Comparing with experiments on
$\mathrm{NiGa_2S_4}$, we argue that the $f$-wave pairing spin liquid
is a plausible ground state.

 It should be noted that rather different fermionic/bosonic spinon approaches have
also been applied to $SU(N)$ spin models where the dynamic symmetry
is generalized from $SU(2)$ to $SU(N)$, and the Hamiltonian is
composed of $SU(N)$ generators.\cite{Affleck&Marston,MF,
Read&Sachdev}. In these approaches, $N$-species of particles are
introduced to construct the $SU(N)$-spin operator and
Hamiltonians.\cite{Read&Sachdev} The major difference between our
approach and the $SU(N)$ approaches is that $(2S+1)$ species of
fermions are introduced to form a $SU(2)$-spin (or irreducible
representation of $SU(2)$ group) in our approach. Unlike the $SU(N)$
models,\cite{Read&Sachdev} our approach is not limited to bipartite
lattices.

The paper is organized as follows. In section II, we introduce the
general fermionic spinon representation for arbitrary spin $S$ and
discuss the symmetry properties associated with the representation.
In section III, we introduce the path integral formalism where
different ways of imposing particle number constraints are
presented. A particle-hole symmetric way of imposing constraint is
discussed. In section IV.A and IV.B, we focus on 1D Heisenberg
models for spin-1 and spin-3/2, respectively, and study two versions
of the mean field theories. The cases for general spin $S$ are
discussed in section IV.C. In section V we study 2D AFM models on
triangular lattice. Our paper is summarized and concluded in section
VI.

\section{General fermionic spin representation}

We begin with the fermionic representation for spins. For spin
$S=1/2$, two species of fermions $c_\uparrow$ and $c_\downarrow$
are introduced to construct the three spin operators $S^{x,y,z}$.
To generalize this fermionic representation to arbitrary spin-$S$,
we introduce $2S+1$ species of fermionic operators $c_m$
satisfying anti-commutation relations,
\begin{eqnarray}\label{Fermi}
\{c_m,c^\dagger_n\}=\delta_{mn},
\end{eqnarray}
where $m,n=S,S-1,\cdots,-S$. The spin operator can be expressed in
terms of $c_m$ and $c^\dagger_n$'s,
\begin{eqnarray*}
\hat {\mathbf S}=C^\dagger {\mathbf I}C,
\end{eqnarray*}
where $C=(c_S, c_{S-1},\cdots,c_{-S})^T$ and $I^a$ $(a=x,y,z)$ is
a $(2S+1)\times(2S+1)$ matrix whose matrix elements are given by
\begin{equation*}
I^a_{mn}=\langle S,m|S^a|S,n\rangle.
\end{equation*}
It is easy to see that the operators $\hat S^a$ satisfy the
$SU(2)$ angular momentum algebra, $[\hat S^a, \hat
S^b]=i\epsilon^{abc}\hat S^c$. Under a rotational operation, $C$
is a spin-$S$ ``spinor" transforming as $C_m\to D^{S}_{mn}C_{n}$
and $\hat{\mathbf S}$ is a vector transforming as $S^a\to
R_{ab}S^{b}$, here $D^S$ is the $2S+1$-dimensional irreducible
representation of $SU(2)$ group generated by $\mathbf I$ and $R$
is the adjoint representation.

As in the $S=1/2$ case, a constraint that there is only one
fermion per site is needed to project the fermionic system into
the proper Hilbert space representing spins, i.e.
\begin{equation}
\label{contr1} (\hat N_i-N_f)|\mathrm{phy}\rangle=0,
\end{equation}
where $i$ is the site index and $N_f=1$ (the particle
representation, one fermion per site). Alternatively, it is
straightforward to show that the constraint $N_f=2S$ (the hole
representation, a single hole per site) represents a spin equally.
The $N_f=1$ representation can be mapped to the $N_f=2S$
representation {\em via} a {\em particle-hole transformation}. For
$S=1/2$, the particle picture and the hole picture are identical,
reflecting an intrinsic particle-hole symmetry of the underlying
Hilbert space which is absent for $S\ge1$.

Following Affleck {\it et al.} ,\cite{AZHA-1/2} we introduce
another ``spinor" $\bar C=(c^\dagger_{-S}, -c^\dagger_{-S+1},
c^\dagger_{-S+2},\cdots,(-1)^{2S}c^\dagger_{S})^T$, whose
components can be written as $\bar C_m=(-1)^{S-m}c^\dagger_{-m}$,
where the index $m$ runs from $S$ to $-S$ as in $C$. To examine
whether $\bar C$ is really a ``spinor", we construct a spin
singlet state for two spins at site $i$ and $j$,
$\frac{C_i^\dagger\bar C_j}{\sqrt{2S+1}} |\mathrm{vac}\rangle
=\frac{1}{\sqrt{2S+1}}\sum_{m=-S}^{S}(-1)^{S-m}|S,m\rangle_i|S,-m\rangle_j$.
Therefore $C^\dagger_i\bar C_j$ is a scalar operator as
$C^\dagger_i C_j$, meaning that $\bar C$ and $C$ must behave
identically under spin rotation. Consequently, the spin operators
can also be written in terms of $\bar C$,
\begin{eqnarray*}
\hat {\mathbf S}=\bar C^\dagger {\mathbf I}\bar C.
\end{eqnarray*}
Combining $C$ and $\bar C$ into a $(2S+1)\times2$ matrix
$\psi=(C,\bar C)$, \cite{AZHA-1/2} we can reexpress the spin
operators as
\begin{eqnarray}\label{spin}
\hat {\mathbf S}=\frac{1}{2}\mathrm{Tr}(\psi^\dagger {\mathbf
I}\psi)
\end{eqnarray}
and the constraint can be expressed as
\begin{eqnarray}\label{contr2}
\label{newcontr1}\mathrm{Tr}(\psi\sigma_z\psi^\dagger)=2S+1-2N_f=\pm(2S-1),
\end{eqnarray}
where the $+$ sign implies $N_f=1$ and $-$ sign implies $N_f=2S$.

We are interested in two kinds of operations (or groups) acting on
$\psi$. The first kind belongs to a $(2S+1)$-dimensional
irreducible representation of $SU(2)$ group, acting on the left of
$\psi$. Suppose $G$ is an element of this irreducible
representation, then $G^\dagger I^aG=R^{ab}I^b$ ($R$ belongs to
the adjoint irreducible representation of $SU(2)$ group, which is
a $SO(3)$ matrix). Under transformation $\psi\rightarrow G\psi$,
the spin operator $\hat S^a=\frac{1}{2}\mathrm{Tr}(\psi^\dagger
I^a\psi)$ becomes $\hat S^a\rightarrow R^{ab}\hat S^b$, which
means \textit{a rotation of the spin}. It is obvious that the
particle number constraint Eq.~(\ref{contr2}) remains unchanged
under the action of $G$.

The other kind belongs to a $2\times2$ unitary group acting on the
right of $\psi$ which keeps the spin operator Eq.~(\ref{spin}) and
the fermionic statistics Eq.~(\ref{Fermi}) invariant. This group
reflects the symmetry properties of the underlying Hilbert space
structure in the fermionic representation. We call it an
\textit{internal symmetry group}. The internal symmetry group is
different for integer and half-integer spins. It is
$U(1)\bar\otimes Z_2=\{e^{i \sigma_z\theta}, \sigma_x
e^{i\sigma_z\theta}= e^{-i\sigma_z\theta}\sigma_x;\theta\in
\mathbb{R}\}$ for the former and $SU(2)$ for the latter. We leave
the rigorous proof in appendix A, and shall explain qualitatively
the reason behind here. Notice that $C$ and $\bar C$ are not
independent. The operators in the internal symmetry group will
``mix" the two fermion operators in the same row of $C$ and $\bar
C$, i.e. $c_{S}$ and $c_{-S}^\dagger$, $c_{S-1}$ and
$-c_{-S+1}^\dagger$, etc. For integer spins, $c_0$ and
$(-1)^Sc_0^\dagger$ will be ``mixed". To keep the relation
$\{c_0,c_0^\dagger\}=1$ invariant, there are only two methods of
``mixing": one is an $U(1)$ transformation, the other is
interchanging the two operators. These operations form the
$U(1)\bar\otimes Z_2$ group. For half-odd-integer spins, the pair
$(c_0,(-1)^Sc_0^\dagger)$ do not exist, and the symmetry group is
the maximum group $SU(2)$. The difference between integer and
half-integer spins is a fundamental property of the fermionic
representation as we shall see more later.

Now let us see how the constraints Eq.~(\ref{newcontr1}) transform
under the symmetry group. For $S=1/2$, the constraint
Eq.~(\ref{newcontr1}) is invariant under the transformation
$\psi\to\psi W$ because the right hand side vanishes (due to the
particle-hole symmetry of the Hilbert space). For integer spins,
if $W=e^{i\sigma_z\theta}$, then $W\sigma_z W^\dagger=\sigma_z$,
and Eq.~(\ref{newcontr1}) is invariant. If
$W=\sigma_xe^{i\sigma_z\theta}$, then $W\sigma_z
W^\dagger=-\sigma_z$, meaning that the ``particle" picture($+$
sign in Eq.~(\ref{contr2})) and the ``hole" picture ($-$ sign in
Eq.~(\ref{contr2})) are transformed to each other.

For a half-odd-integer spin with $S\ge 3/2$, $W\in SU(2)$ is a
rotation and we need to extend the constraint into a vector form
similar to $S=1/2$ case,\cite{AZHA-1/2} so that Eq.~(\ref{contr2})
becomes
\begin{eqnarray}\label{veccontr}
\mathrm{Tr}(\psi\pmb\sigma\psi^\dagger)=(0,0,\pm(2S-1))^T.
\end{eqnarray}
Under the group transformation $\psi\to\psi W$,
\begin{eqnarray}\label{veccontr2}
\mathrm{Tr}(\psi\pmb\sigma\psi^\dagger)\rightarrow(R^{-1})(0,0,\pm(2S-1))^T.
\end{eqnarray}
where $W\sigma^aW^\dagger=R_{ab}\sigma^b$, $a,b=x,y,z$, i.e. $R$
is a 3 by 3 matrix representing a 3D rotation in the internal
Hilbert space. The transformed constraint represents a new Hilbert
subspace which is still a $(2N+1)$-dimensional irreducible
representation of the spin $SU(2)$ algebra. Any measurable
physical quantity such as the spin $\mathbf S$ remains unchanged
in the new Hilbert space. Therefore, for half-odd-integer spins
($S\ge 3/2$), there exists infinitely many ways of imposing the
constraint that gives rise to a Hilbert subspace representing a
spin. However, for integer spins, there exists only two possible
constraint representations.

 The different representations of constraints for $S>1/2$ systems
result in different path integral representations and different
mean field theories. These mean field theories are equivalent in
the sense that they can be transformed to each other by the
internal symmetry group. In the next section, we shall study the
Heisenberg model in different representations and shall construct
a ``mixed" Path Integral representation which restored
particle-hole symmetry, and the internal symmetry group becomes
``almost" a gauge symmetry.\cite{AZHA-1/2} The ``mixed"
representation is studied in section IV where a new mean field
theory is proposed which recovers Haldane conjecture at 1D for the
Heisenberg model.

\section{Path integral formalism and mean field theory for Heisenberg model}

 We shall focus on the antiferromagnetic Heisenberg model
$H=J\sum_{\langle i,j\rangle}\mathbf {S}_i\cdot\mathbf{S}_j$ with
$J>0$ in the rest of the paper. We start by presenting some useful
algebraic manipulations of the Hamiltonian, and then discuss the
general path integral formalism. The different ways of handling
constraints and how they affect the symmetry of the Lagrangian
will be discussed in the process.

\subsection{Heisenberg model and an effective Hamiltonian}

It is known that for spin-1/2 the Heisenberg interaction can be
written as \cite{AZHA-1/2}
\begin{eqnarray}\label{S1/2}
\hat{\mathbf{S}}_i\cdot\hat {\mathbf{S}}_j&=&-\frac{1}{8}
\mathrm{Tr}:(\psi_i^\dagger\psi_j \psi_j^\dagger\psi_i):\nonumber
\\&=&-\frac{1}{4}:(\chi_{ij}^\dagger\chi_{ij}+\Delta_{ij}^\dagger\Delta_{ij}):,
\end{eqnarray}
where
\begin{equation}
\chi_{ij}=C_i^\dagger C_j, ~\quad \Delta_{ij}=\bar C_i^\dagger
C_j.
\end{equation}
are spin-singlet operators. $\chi_{ij}$ and $\Delta_{ij}$ will be
extended to arbitrary spin magnitudes with the above definition
and will be used in the following discussions. Interestingly, an
expression almost the same as Eq.~(\ref{S1/2}) holds for $S=1$,
but this is no longer true for larger spins. We shall present
precise formula for $S=3/2$ and $S=2$, and provide a general
discussion for higher spins.

For $S=1$, the three spin matrices are
\begin{eqnarray*}
I_+=(I_-)^\dagger=\left(\begin{matrix}0 & \sqrt2 & 0\\ 0 & 0 & \sqrt2\\
0 & 0 & 0 \end{matrix}\right), I_z=\left(\begin{matrix}1 & 0 & 0\\
0 & 0 & 0\\ 0 & 0&-1\end{matrix}\right).
\end{eqnarray*}
where $I_{\pm}=I_x\pm iI_y$. The matrix operator $\psi$ is given
by
\[ \psi=(C\ \bar C)=\left(\begin{matrix}c_1&c^\dagger_{-1}
\\c_0&-c_0^\dagger \\c_{-1}&c^\dagger_1\end{matrix}\right).  \]
The spin operator can be written in form of Eq.~(\ref{spin}), and
it can be shown after some straightforward algebra that the
Hamiltonian can be written as
\begin{eqnarray}\label{H1}
H&=&J\sum_{\langle i,j\rangle} \hat{\mathbf{S}}_i\cdot\hat
{\mathbf{S}}_j=-\frac{J}{2}\sum_{\langle i,j\rangle}
\mathrm{Tr}:(\psi_i^\dagger\psi_j\psi_j^\dagger\psi_i):\nonumber\\
&=&-J\sum_{\langle i,j\rangle}:(\chi_{ij}^\dagger\chi_{ij}
+\Delta_{ij}^\dagger\Delta_{ij}):.
\end{eqnarray}

For $S=3/2$, the three spin matrices are
\begin{eqnarray*}
I_+=(I_-)^\dagger=\left(\begin{matrix}0 & \sqrt3 & 0 & 0\\ 0 & 0 & 2 & 0\\
0 & 0 & 0 & \sqrt3\\0&0&0&0\end{matrix}\right),
I_z=\left(\begin{matrix}\frac{3}{2} & 0 & 0 &0\\ 0 & \frac{1}{2} & 0 &0\\
0 & 0&-\frac{1}{2}&0\\0&0&0&-\frac{3}{2}\end{matrix}\right).
\end{eqnarray*}
and the matrix operator $\psi$ is given by $\psi=(C\ \bar C)$
where
$C=(c_{{3\over2}},c_{{1\over2}},c_{-{1\over2}},c_{-{3\over2}})^T$
and $\bar C= (c^\dagger_{-{3\over2}},-c_{-{1\over2}}^\dagger,
c^\dagger_{{1\over2}},-c_{{3\over2}}^\dagger)^T$. In this case the
singlet operators $\chi_{ij}$ and $\Delta_{ij}$ alone are not
enough to represent the Heisenberg Hamiltonian, and triplet hoping
and pairing terms are necessary. After some straightforward but
tedious algebra, we find that the Hamiltonian for spin $S=3/2$
Heisenberg model can be written as
\begin{eqnarray}\label{H3/2}
H&=&J\sum_{i,j} \hat{\mathbf{S}}_i\cdot\hat{\mathbf{S}}_j\nonumber\\
&=&-\frac{J}{4}\sum_{i,j}:[\mathrm{Tr}(\psi_i^\dagger\mathbf{I}
\psi_j\cdot\psi_j^\dagger\mathbf{I}\psi_i)+S^2\mathrm{Tr}
(\psi_i^\dagger\psi_j\psi_j^\dagger\psi_i)]:\nonumber\\
\end{eqnarray}
It is interesting to note that the above form holds also for $S=1$
\cite{S=1/2}. However, for $S=1$, the two terms in the square
bracket can be transformed from one to another, which does not
hold for $S=3/2$. This suggests that $S=1/2,1$ Heisenberg models
have a higher ``hidden symmetry" compared with $S=3/2$. Actually,
there is another hidden symmetry for the spin $S=1/2$ systems
which is absent for spin-1,
\[ :\chi_{ij}^\dagger\chi_{ij}: =:\Delta_{ij}^\dagger\Delta_{ij}:-\hat N_i\hat N_j,  \]
where $\hat N_i=C_i^\dagger C_i$ is the particle number on site
$i$. Our analysis suggests that smaller spins have higher symmetry
when the spin interaction is expressed in the fermionic
representation.

For $S$=2, the Heisenberg Hamiltonian can be written as
\[H=J\sum_{\langle i,j\rangle} \hat{\mathbf{S}}_i\cdot\hat
{\mathbf{S}}_j=-\frac{J}{2}\sum_{\langle i,j\rangle}
\mathrm{Tr}:(\psi_i^\dagger\mathbf{I}\psi_j\cdot
\psi_j^\dagger\mathbf{I}\psi_i):.\] As in the case of lower spins,
the terms $\mathrm{Tr}:(\psi_i^\dagger\mathbf{I}\psi_j\cdot
\psi_j^\dagger\mathbf{I}\psi_i):$ and
$\mathrm{Tr}:(\psi_i^\dagger\psi_j\psi_j^\dagger\psi_i):$ have
finite overlaps and are not completely independent of each other.
For $S>2$, it is not possible to represent the Hamiltonian in the
above two terms alone. Quintet and higher multipolar hoping and
pairing operators are needed to represent the Heisenberg
Hamiltonian and we are not able to obtain any general expression.

Summarizing, the Heisenberg model can be written in the fermionic
representation as
\begin{eqnarray}\label{Htr}
H &=&-{J\over2}\sum_{\langle
i,j\rangle}\left[a(S)\mathrm{Tr}:(\psi_i^\dagger\psi_j
\psi_j^\dagger\psi_i):\right.\nonumber\\&&\left.+b(S)\mathrm{Tr}:(\psi_i^\dagger\mathbf{I}\psi_j\cdot
\psi_j^\dagger\mathbf{I}\psi_i):+\cdots\right]
\end{eqnarray}
where $a(S), b(S),\cdots$ are parameters dependent on the spin
magnitude $S$ and $\cdots$ represents quintet and higher multiplet
terms.  In the following, we shall study in detail a Hamiltonian
which keeps the first two terms only. The Hamiltonian can be
considered as an effective Hamiltonian for constructing trial
ground state wavefunctions in a variational calculation. Notice
that the effective Hamiltonian is in fact ``exact" up to $S=2$ if
$a(S)$ and $b(S)$ are chosen properly.

\subsection{Path integral formalism and constraints}
In imaginary time path integral formalism, the partition function
is given by
\begin{eqnarray}\label{U}
&&Z=\mathrm{Tr}(e^{-\beta H})=\int D\psi D\psi^\dagger D\lambda
e^{-\int_0^\beta
L(\psi,\psi^\dagger,\lambda)d\tau}\nonumber\\
&&L=\sum_i\frac{1}{2}\left[\mathrm{Tr}\psi_i(\partial_\tau-i\lambda_i\sigma_z)
\psi^\dagger_i\pm i(2S-1)\lambda_i\right]+H\nonumber\\
\end{eqnarray}
where $\lambda_i$ is the Lagrange multiplier field introduced to
impose the constraint\ Eq.~(\ref{contr2}). The $+(-)$ sign
corresponds to the particle ($\hat N=1$) and hole ($\hat N=2S$)
representations. The Grassmann number $\psi$ satisfy the
antiperiodic boundary condition $\psi(0)=-\psi(\beta)$.

Notice that the Lagrangian Eq.~(\ref{U}) is not invariant under
the action of the internal symmetry group because of the
non-invariant particle number constraint Eq.~(\ref{contr2}), i.e.
the internal symmetry group of the spin operator is not a symmetry
group of the Lagrangian. In the following we will employ a trick
to restore the symmetry of the Lagrangian. The integer- and
half-odd-integer- spin models will be discussed separately because
of the intrinsic difference in their internal symmetry groups.

\subsubsection{Integer spins}
Our trick is to consider an average of all the partition functions
with different particle-number constraint representations. In the
integer spin case, the average is given by
\begin{eqnarray}\label{path1}
Z&=&\int D\psi D\psi^\dagger e^{-\int_0^\beta
(\sum_i\frac{1}{2}\mathrm{Tr}[\psi_i\partial_\tau\psi^\dagger_i]+H)d\tau}\times\nonumber\\&&
\prod_{\{i,\tau\}}\frac{1}{2}\left[\delta(\hat N_i-1)+\delta(\hat N_i-2S)\right] \nonumber\\
&=&\int D\psi D\psi^\dagger D\lambda e^{-\int_0^\beta
(\sum_i\frac{1}{2}\mathrm{Tr}[\psi_i\partial_\tau\psi^\dagger_i]+H)d\tau}\times\nonumber\\&&
\prod_{\{i,\tau\}}\frac{1}{2}\left(e^{i\frac{\lambda_i}{2}[\mathrm{Tr}(\psi_i\sigma_z
\psi_i^\dagger)-2S+1]}\right.\nonumber\\&&\left.
+e^{i\frac{\lambda_i}{2}[\mathrm{Tr}(\psi_i\sigma_z\psi_i^\dagger)+2S-1]}\right).
\end{eqnarray}
Notice that we do not make any approximations in deriving
Eq.~(\ref{path1}) from Eq.~(\ref{U}) and the ``new" partition
function is a faithful representation of Heisenberg model except
that it averages over all possible ``particle" and ``hole"
representations of constraints locally. Similar ideas have been
applied to generate the supersymmetric representation\cite{cheng&Ng}
and the $SU(2)$ representation of the $t$-$J$ model \cite{g4}. The
fermion particle-hole symmetry is restored in this representation.
The Lagrangian corresponding to Eq.~(\ref{path1}) is
\begin{eqnarray}
L&=&\frac{1}{2}\sum_i\left[\mathrm{Tr}[\psi_i(\partial_\tau-i\lambda_i\sigma_z)\psi^\dagger_i]
-2\ln\cos\frac{(2S-1)\lambda_i}{2}\right]\nonumber\\
&&+H.\nonumber
\end{eqnarray}
This form of Lagrangian is invariant under the internal symmetry
group $U(1)\bar\otimes Z_2$ and the symmetry group becomes almost
a ``gauge symmetry" of the new Lagrangian as we shall see in the
following. We note that the Lagrangian is complex because of the
multiplier $\lambda$. This problem can be solved by lifting the
contour of integration over $\lambda$ into complex plane via
analytic continuation to find a saddle point in the imaginary axis
\cite{LeeLee}. For convenience, we write $\tilde
\lambda_i=i\lambda_i$, and integrate $\tilde\lambda_i$ from
$-i\infty$ to $i\infty$ (the saddle point of $\tilde\lambda_i$ is
real). Then the Lagrangian becomes
\begin{eqnarray}\label{L_GI}
L&=&\frac{1}{2}\sum_i\left[\mathrm{Tr}[\psi_i(\partial_\tau-\tilde\lambda_i\sigma_z)\psi^\dagger_i]
-2\ln\cosh\frac{(2S-1)\tilde\lambda_i}{2}\right]\nonumber\\ &&+H,
\end{eqnarray}

We shall now consider the path integral representation in terms of
the effective Hamiltonian Eq.~(\ref{Htr}) keeping only the first
two terms. The effective Hamiltonian can be decoupled by a
Hubbard-Stratonovich matrix field (see Appendix B)
\begin{eqnarray*}
\hat U_{ij}&=&\psi_i^\dagger\psi_j=\left(\begin{matrix}\chi_{ij}
&-\Delta_{ij}^\dagger\\ \Delta_{ij}&-\chi_{ij}^\dagger\end{matrix}\right),\\
\hat{\mathbf V}_{ij}&=&\psi_i^\dagger\mathbf I\psi_j= \left(\begin{matrix}\mathbf v_{ij} & \mathbf u_{ij}^\dagger\\
\mathbf u_{ij} & \mathbf v_{ij}^\dagger\end{matrix}\right),
\end{eqnarray*}
where $\mathbf u_{ij}=C_i^\dagger{\mathbf I}C_j$, and $\mathbf
v_{ij}=\bar C_i^\dagger{\mathbf I}C_j$. Note that $\mathbf u_{ij}$
and $\mathbf v_{ij}$ form two sets of spin triplet operators,
respectively. The Hamiltonian becomes
\begin{eqnarray}\label{HB}
H_{}&=&{J\over2}\sum_{\langle
i,j\rangle}\left\{a(S)\mathrm{Tr}[\hat U_{ij}^\dagger\hat
U_{ij}-(\hat U_{ij}^\dagger\psi_i^\dagger\psi_j+h.c.)]\right.\nonumber\\
&&\left.+b(S)\mathrm{Tr}[\hat {\mathbf V}_{ij}^\dagger\cdot\hat
{\mathbf V}_{ij} -(\hat{\mathbf V}_{ij}\cdot \psi_j^\dagger\mathbf
I\psi_i+h.c.)]\right\}.
\end{eqnarray}
The $\hat U_{ij}, \hat{\mathbf V}_{ij}$ and $\tilde\lambda_i$
fields define $U(1)\bar\otimes Z_2$ lattice gauge fields coupling
to the fermionic particles which we shall call spinons in the
following. Under the local gauge transformation
$\psi_i\rightarrow\psi_i W_i$, the temporal component and the
spacial component of the gauge fields transform as
\begin{eqnarray}\label{GaugeTrans}
\tilde\lambda_i\sigma_z&\rightarrow& W_i^\dagger(\tilde\lambda_i\sigma_z-\frac{d}{d\tau})W_i\nonumber,\\
\hat U_{ij}&\rightarrow& W_i^\dagger\hat U_{ij} W_j\nonumber,\\
\hat {\mathbf V}_{ij}&\rightarrow& W_i^\dagger\hat {\mathbf
V}_{ij} W_j.
\end{eqnarray}
Notice that $\lambda_i$ changes sign under an uniform $Z_2$ gauge
transformation, and $\chi_{ij}$, $\Delta_{ij}$ exchange their
roles under a ``staggered" $Z_2$ transformation where we have
$\sigma_x$ on one sublattice and $\mathbb{I}$ on the other
sublattice. Notice also that the Lagrangian is not invariant under
a time-dependent gauge transformation, because the term
$2\ln\cosh\frac{(2S-1)\tilde\lambda_i}{2}$ is not invariant. A
similar situation occurs in the supersymmetric representation of
Heisenberg model \cite{cheng&Ng}. Because of this restriction the
internal symmetry group does not generate a complete ``gauge
symmetry" in the path integral formalism.

Integrating out the fermion fields, we get an effective action in
terms of the $U_{ij}$, $\hat{\mathbf V}_{ij}$ and $\lambda_i$
fields. The mean field values of the these fields are given by the
saddle point of the effective action determined by the
self-consistent equations $U=\langle \hat U_{ij}\rangle$, $\mathbf
V=\langle \hat {\mathbf V}_{ij}\rangle$ with the mean field
Hamiltonian
\begin{eqnarray}\label{Heff1}
H_{m} &=&J\sum_{\langle i,j\rangle}\left[a(S)[-(\chi^* C_i^\dagger
C_j+\Delta^*\bar C_i^\dagger C_j+h.c.)+|\chi|^2\right.\nonumber\\
&&\left.+|\Delta|^2]+b(S)[-(\mathbf v^*\cdot C_i^\dagger\mathbf I
C_j+\mathbf u^*\cdot\bar C_i^\dagger\mathbf I C_j+h.c.)\right.\nonumber\\
&&\left.+|\mathbf u|^2+|\mathbf v|^2]\right]
+\sum_i[\tilde\lambda(\hat N_i-\frac{2S+1}{2})\nonumber\\&&
-\ln\cosh\frac{(2S-1)\tilde\lambda}{2}],
\end{eqnarray}
where $\chi=\langle C_i^\dagger C_j\rangle$, $\Delta=\langle \bar
C_i^\dagger C_j\rangle$, $\mathbf u=\langle \bar C_i^\dagger \mathbf
I C_j\rangle$, $\mathbf v=\langle C_i^\dagger \mathbf I C_j\rangle$,
and $\tilde\lambda$ is determined by the condition \[ \langle\hat
N_i\rangle-{2S+1\over2}={2S-1\over2}\tanh{(2S-1)\tilde\lambda\over2}.
\]
Notice that the averaged particle number satisfies $1<\langle\hat
N_i\rangle<2S$ because we are mixing the particle and hole
pictures in the constraint. If the ground state doesn't break the
particle-hole symmetry, then $\tilde\lambda=0$ and we get the
half-filling condition $\langle\hat N_i\rangle=(2S+1)/2$.

\subsubsection{Half odd integer spins}
We first examine how the two particle number constrains
$\mathrm{Tr}(\psi\sigma_z\psi^\dagger)=\pm(2S-1)$ are transformed
into each other by the $SU(2)$ internal symmetry group. We
consider the vector constraint Eq.~(\ref{veccontr}) for $S\ge1/2$
\begin{equation*}
\hat{\mathbf M}=\frac{1}{2S-1} \mathrm{Tr}(\psi\pmb\sigma
\psi^\dagger)=(0,0,1)^T,
\end{equation*}
where
\begin{eqnarray*}
&&\hat M_x=\frac{1}{2S-1}\mathrm{Tr}(\bar CC^\dagger+C\bar
C^\dagger),\nonumber\\
&&\hat M_y=\frac{i}{2S-1}\mathrm{Tr}(\bar CC^\dagger-C\bar
C^\dagger),\nonumber\\
&&\hat M_z=\frac{1}{2S-1}\mathrm{Tr}(CC^\dagger-\bar C\bar
C^\dagger).
\end{eqnarray*}

The $M_x$ and $M_y$ components indicate that the one-site pairings
$c_{{S}}c_{-{S}}, \cdots,$  $c_{{1\over2}}c_{-{1\over2}}$ are
equal to 0, and the $M_z$ component imposes the constraint that
there is exactly one fermion per site. Notice that the first two
components of the constraint is automatically satisfied if the
third component is satisfied rigorously. Under the $SU(2)$
transformation $\psi\rightarrow \psi W$, the constraint becomes
$\frac{1}{2S-1}\mathrm{Tr}(\psi\pmb\sigma\psi^\dagger)=R^{-1}
(0,0,1)^T$, where $W\sigma_a W^\dagger=R_{ab} \sigma_b$, $R$ is an
$SO(3)$ rotation. Since the constraint is invariant under an
$U(1)$ gauge transformation generated by $\sigma_z$, the continuum
space formed by different constrains is the surface of a sphere
$SU(2)/U(1)=S^2$. In particular, for $W=e^{i\sigma_x\pi/2}$, the
north pole of the sphere which corresponds to the positive sign in
Eq.~(\ref{contr2}) is transformed to the south pole,
$R^{-1}(0,0,1)^T=(0,0,-1)^T$, corresponding to the negative sign
in Eq.~(\ref{contr2}).

Now we construct a Lagrangian which is invariant under the $SU(2)$
internal symmetry group. In the path integral formalism, the
particle number constraint
$\frac{1}{2S-1}\mathrm{Tr}(\psi_i\pmb\sigma\psi^\dagger_i)= (\hat
M_{i}^x,\hat M_{i}^y,\hat M_{i}^z)^T=(0,0,1)^T$ can be realized by
introducing a vector Langrange multiplier field $\pmb{\lambda}_i$
\begin{eqnarray}
\delta(\hat M_{i}^x)\delta(\hat M_{i}^y)\delta(\hat
M_{i}^z-1)=(\frac{1}{2\pi})^3\int d\lambda_{i}^xd\lambda_{i}^y
d\lambda_{i}^z\nonumber\\ \exp\{i[\lambda_{i}^x\hat
M_{i}^x+\lambda_{i}^y\hat M_y+\lambda_i^z(\hat M_{i}^z-1)]\}.
\end{eqnarray}
The vector $(0,0,1)^T$ changes into
$R^{-1}(0,0,1)^T=(n^x,n^y,n^z)^T=\hat{\mathbf n}$ under $SU(2)$
gauge transformation, where $\hat{ \mathbf n}$ is an unit vector
on the surface of the sphere $S^2$. As in the integer spin case, a
$SU(2)$ invariant path integral formalism can be obtained by
averaging over all possible particle number constraints:
\begin{eqnarray}
&&\langle\delta(\hat M_{i}^x-n^x)\delta(\hat
M_{i}^y-n^y)\delta(\hat M_{i}^z-n^z)\rangle_{ \{\hat n\}} \nonumber\\
&=&\frac{1}{4\pi}(\frac{1}{2\pi})^3\int d^2\hat{ \mathbf n}\int
d^3\lambda_{i}\exp\{i\pmb
{\lambda_i}\cdot(\hat{\mathbf M}_i- \hat{ \mathbf n})\}\nonumber\\
&=&(\frac{1}{2\pi})^3\int d^3\lambda_i \exp\{i(\pmb
{\lambda}_i\cdot\hat{\mathbf
M}_i-i\ln\frac{\sin\lambda_i}{\lambda_i})\}
\end{eqnarray}
where
$\lambda_i=\sqrt{(\lambda_i^{x})^2+(\lambda_i^{y})^2+(\lambda_i^{z})^2}$.
Introducing $\pmb{\tilde\lambda}_i=i\pmb\lambda_i$, we obtain a
Lagrangian with ``almost" $SU(2)$ gauge symmetry:
\begin{eqnarray}\label{LSU2}
L&=&\frac{1}{2}\sum_i[\mathrm{Tr}[\psi_i(\partial_\tau-\pmb{\tilde\lambda}_i\cdot\pmb\sigma)\psi^\dagger_i]
-(2S-1)\ln\frac{\sinh\tilde\lambda_i}{\tilde\lambda_i}]\nonumber\\&&+H.
\end{eqnarray}

As in the case of integer spins, we replace $H$ by the effective
Hamiltonian Eq.~(\ref{Htr}) with only the first two terms.
Introducing the Hubbard Stratonovich field as in the integer spin
case (see Appendix B for details),
\begin{eqnarray*}
\hat U_{ij}&=&\psi_i^\dagger\psi_j=\left(\begin{matrix}\chi_{ij}&\Delta_{ij}^\dagger\\
\Delta_{ij}&-\chi^\dagger_{ij}\end{matrix}\right),\nonumber\\
\hat{\mathbf V}_{ij}&=&\psi_i^\dagger\mathbf I\psi_j= \left(\begin{matrix}\mathbf v_{ij} & -\mathbf u_{ij}^\dagger\\
\mathbf u_{ij} & \mathbf v_{ij}^\dagger\end{matrix}\right),\nonumber\\
\end{eqnarray*}
then $\hat U_{ij},\hat{\mathbf V}_{ij}$ and
$\pmb{\tilde\lambda}_i$ form an $SU(2)$ lattice gauge field
coupling to the spinons. (Notice the sign difference in the matrix
elements of $U$ and $\hat{\mathbf V}$ between integer and
half-odd-integer spins.) Under gauge transformation
$\psi_i\rightarrow\psi_i W_i$ ($W_i\in SU(2)$) the gauge fields
transform as,
\begin{eqnarray*}
\pmb{\tilde\lambda}_i\cdot\pmb\sigma&\rightarrow& W_i^\dagger(\pmb
{\tilde\lambda}_i\cdot\pmb\sigma-\frac{d}{d\tau})W_i\nonumber\\
\hat U_{ij}&\rightarrow& W_i^\dagger\hat U_{ij} W_j\nonumber\\
\hat {\mathbf V}_{ij}&\rightarrow& W_i^\dagger\hat {\mathbf
V}_{ij} W_j,
\end{eqnarray*}
and the Lagrangian Eq.~(\ref{LSU2}) is invariant only under
time-independent gauge transformations as in the integer spin case
because of the non-invariant term
$(2S-1)\ln\frac{\sinh\tilde\lambda_i}{\tilde\lambda_i}$. This is
different from the $S=1/2$ case, where the Lagrangian is invariant
under time-dependent gauge transformations. The mean field
effective Hamiltonian corresponding to Eq.~(\ref{LSU2}) is
\begin{eqnarray}\label{Heff2}
H_{m} &=&J\sum_{\langle i,j\rangle}\left[a(S)[-(\chi^* C_i^\dagger
C_j+\Delta^*\bar C_i^\dagger C_j+h.c.)+|\chi|^2\right.\nonumber\\
&&\left.+|\Delta|^2]+b(S)[-(\mathbf v^*\cdot C_i^\dagger\mathbf I
C_j+\mathbf u^*\cdot\bar C_i^\dagger\mathbf I C_j+h.c.)\right.\nonumber\\
&&\left.+|\mathbf u|^2+|\mathbf v|^2]\right]
-\sum_i{2S-1\over2}(\pmb{\tilde\lambda}\cdot\hat {\mathbf M}_i
+\ln\frac{\sinh\tilde\lambda}{\tilde\lambda})\nonumber\\
\end{eqnarray}
where the parameters $\chi, \Delta, \mathbf u,\mathbf v$ are
solved self-consistently by the mean field equations $U=\langle
\hat U_{ij}\rangle$, $\mathbf V=\langle \hat {\mathbf
V}_{ij}\rangle$ and $\pmb{\tilde\lambda}$ is determined by
\[
 \langle\hat{\mathbf M}_i\rangle+{\pmb{\tilde\lambda}\over\tilde\lambda^2}(\tilde\lambda\coth\tilde\lambda-1)=0. \]
 The averaged particle number per site is again given by $\langle\hat N_i\rangle=(2S+1)/2$ if the
ground state respects particle-hole symmetry
$(\pmb{\tilde\lambda}=0)$.

\subsection{An important difference between integer and half-odd-integer spins}\label{sec:to prove}

It can be proved (see Appendix B) that for integer spins,
$\chi_{ji}=\chi_{ij}^\dagger$, $\Delta_{ji}=-\Delta_{ij}$,
$\mathbf u_{ji}=\mathbf u_{ij}$ and $\mathbf v_{ji}=\mathbf
v_{ij}^\dagger$, whereas for half-odd-integer spins,
$\chi_{ji}=\chi_{ij}^\dagger$, $\Delta_{ji}=\Delta_{ij}$, $\mathbf
u_{ji}=-\mathbf u_{ij}$ and $\mathbf v_{ji}=\mathbf
v_{ij}^\dagger$. Notice that the parity of the paring terms
$\Delta_{ij}$ and $\mathbf u_{ij}$ are different for integer and
half-odd-integer spins. These operators are central to the mean
field theory, because the ground states are completely determined
by their expectation values. In particular, for a mean field
theory with $\Delta,\chi\neq0$ and $\mathbf u, \mathbf v=0$, the
mean field ground state is a BCS spin-singlet pairing state where
the order parameter $\langle\Delta_{ij}\rangle$ has even parity
for half-odd-integer spins, but has odd parity for integer spins.
We shall see how this important difference leads to different
excitation spectrums between integer and half-odd-integer spin
systems in the fermionic mean field theory. Similar result exists
for states with $\Delta,\chi=0$ and $\mathbf u, \mathbf v\neq0$.
In this case the mean field ground state is a BCS spin-triplet
pairing state and the parity of the order parameters are reversed.

We note that as in the $S=1/2$ case, the mean field Hamiltonian
should be viewed as a trial Hamiltonian for the ground state
wavefunction of the spin systems after Gutzwiller projection. For
the mean field theory with particle-hole symmetry the Gutzwiller
projection is rather non-trivial. The state is a coherent
superposition of states which allows sites with both one fermion
and with $2S$ fermions. The numerical analysis of such a state is
complicated and we shall not go into details in this paper.

\section{Fermionic mean field theory in 1D and Haldane conjecture}

In this section we apply our mean field theory to the
antiferromagnetic Heisenberg model in one dimension. We first
consider the cases of spin $S=1$ and $S=3/2$ where two versions of
mean field theories based on different methods of implementing the
constraints will be discussed. The mean field results are
summarized in section IV.C where the case of general spin $S$ and
Haldane conjecture will be discussed.

To simplify our analysis we make use of the Wagner-Mermin theorem
which asserts that a continuous symmetry cannot spontaneously break
in one dimension\cite{MSpinWave}. Therefore we assume in our
mean-field theory that the ground state is a spin singlet (spin
liquid state) and the expectation values of $\langle\mathbf
u_{ij}\rangle$ and $\langle\mathbf v_{ij}\rangle$ are zero, since
the rotational symmetry will be broken otherwise. As a result, we
keep only the $a(S)$ term in the trial Hamiltonian. For simplicity
we shall also restrict ourselves to translational invariant
solutions of the mean-field theory in this paper.

\subsection{integer spin: $S=1$}

We choose $H_{tr}$ to be the same as Eq.~(\ref{H1}), i.e.,
$a(1)=-1, b(1)=0$. For convenience of discussion, we introduce two
new fermions $c_s, c_a$ which are the symmetric and antisymmetric
combinations of $c_1$ and $c_{-1}$,
\begin{eqnarray}
c_{si}&=&(c_{1i}+c_{-1i})/\sqrt2\nonumber\\
c_{ai}&=&(c_{1i}-c_{-1i})/\sqrt2
\end{eqnarray}
The mean field Hamiltonian Eq.~(\ref{Heff1}) becomes completely
decoupled in terms of $c_s, c_a$ and $c_0$. In Fourier space, it
can be written as
\begin{eqnarray}\label{Hm1}
H_{m}&=&\sum_k\chi_k(c^\dagger_{sk}c_{sk} +c_{ak}^\dagger c_{ak}+c_{0k}^\dagger c_{0k})\nonumber\\
&&-\sum_k[\frac{\Delta_k^*}{2}(c_{s-k}c_{sk}-c_{a-k}c_{ak}-c_{0-k}c_{0k})+h.c.]\nonumber\\
&&+JN(|\chi|^2+|\Delta|^2)-N(\frac{3}{2}\tilde\lambda+\ln\cosh\frac{\tilde\lambda}{2}),
\end{eqnarray}
where $\chi_k=\tilde\lambda-2J\chi\cos k$,
$\Delta_k^*=2iJ\Delta\sin k$ (since the phases of $\chi$ and
$\Delta$ are unimportant in 1D, we choose $\chi$ and $\Delta$ to
be real numbers) and $N$ is the length of the chain. The mean
field Hamiltonian can be diagonalized by a Bogoliubov
transformation:
\begin{eqnarray}\label{Boglbv_1d}
\gamma_{sk}&=&u_kc_{sk}+v_kc_{s-k}^\dagger,\nonumber\\
\gamma_{ak}&=&u_kc_{ak}-v_kc_{a-k}^\dagger,\nonumber\\
\gamma_{0k}&=&u_kc_{0k}-v_kc_{0-k}^\dagger.
\end{eqnarray}
The coefficients $u_k,v_k$ satisfy the Bogoliubov-de Gennes (BdG)
equations,
\begin{eqnarray}\label{BdG-k}
E_ku_k&=&\chi_ku_k-\Delta_k^*v_k\nonumber\\
E_kv_k&=&-\Delta_ku_k-\chi_kv_k.
\end{eqnarray}

Solving the equations we obtain
\begin{subequations}
\begin{eqnarray}
E_k&=&\sqrt{(\lambda-2J\chi\cos k)^2+(2J\Delta\sin
k)^2}\label{Ex1},\\u_k&=&\cos{\theta_k\over2},\ \ \
v_k=i\sin{\theta_k\over2}\label{ukvk}
\end{eqnarray}
\end{subequations}
where $\theta_k$ is given by $\tan\theta_k={\Delta_k^*\over
i\chi_k}$ and the diagonalized Hamiltonian is
\begin{eqnarray*}
H_m=\sum_{k>0}E_k(\gamma_{sk}^\dagger\gamma_{sk}
+\gamma_{ak}^\dagger\gamma_{ak}+\gamma_{0k}^\dagger\gamma_{0k})+E_0,
\end{eqnarray*}
where $E_0=\sum_k(\chi_k-E_k)+JN(\chi^2+\Delta^2)
-N(\frac{3}{2}\tilde\lambda+\ln\cosh\frac{\tilde\lambda}{2})$ is
the ground state energy. Minimizing $E_0$, we obtain the
self-consistent mean field equations at zero temperature,
\begin{subequations}\label{MFEQ}
\begin{eqnarray}
&&\chi=\frac{2S+1}{2N}\sum_k\cos k(1-\frac{\chi_k}{E_k}),\\
&&\Delta=\frac{2S+1}{2N}\sum_k\sin k\frac{\Delta_k}{E_k},\\
&&\langle\frac{2S+1}{2}-\hat N_i\rangle=\frac{2S+1}{2N}\sum_k
\frac{\chi_k}{E_k}\nonumber\\&&\ \ \ \ \ \ \ \ \ \ \ \ \ \ \ \ \ \
\ \
=-\frac{2S-1}{2}\tanh\frac{(2S-1)\tilde\lambda}{2}.\label{MFEQc}
\end{eqnarray}
\end{subequations}
with $S=1$. The above equations are solved numerically where we
obtain $\chi=\Delta=3/4$ and $\tilde\lambda=0$ as shown in the
first row of Table \ref{tab:spin1}. The averaged particle number
of the spinons per site is $3/2$, indicating that the ground state
respects particle-hole symmetry. The ground state energy is
$E_0=-J(\chi^2 +\Delta^2)$, and the excitation spectrum is flat
(i.e. $k$-independent) with a finite energy gap $3J/2$. The
spin-spin correlation function is
\begin{eqnarray*}
\langle\mathbf S_i\cdot\mathbf S_{i+r}\rangle
&=&\frac{3}{2N^2}\sum_{p,q}e^{i(q-p)r}\left[(1+\frac{\chi_p}{E_p})
(1-\frac{\chi_{q}}{E_{q}})\right.\\&&\left.-\frac{\Delta_p^*}{E_p}\frac{\Delta_{q}}{E_{q}}\right]\\
\end{eqnarray*}
and is nonzero only for $r = 0$ and $r = 1$, with $\langle\mathbf
S_i\cdot \mathbf S_{i}\rangle=3/2$ and $\langle\mathbf S_i\cdot
\mathbf S_{i+1}\rangle=3/4$. The correlation function is zero for
$r\ge2$, indicating that the mean field theory describes a
short-ranged Valence-Bond-Solid (VBS) state.

\begin{table}
\caption{Solutions of two versions of mean fields for spin-1 AF
Heisenberg chain. $E$ is the ground state energy per site and
$E_{gap}$ is the excitation gap.}\label{tab:spin1}
\begin{ruledtabular}
\begin{tabular}{ccccccc}
$\langle N_i\rangle$      &    $\chi$     & $\Delta$      & $\tilde\lambda$ &   $E(J)$       & $E_{gap}(J)$&$\langle\mathbf S_i\cdot\mathbf S_i\rangle$\\
\hline          3/2       & $3/4$         &$3/4$          &     0     &$-{9/8}$        & ${3/2}$     &                   3/2                     \\
                 1        &$0.7625$       &${0.5671}$     &   0.7709  &$-{0.9032}$     & ${0.3634}$  &                   4/3                     \\
\end{tabular}
\end{ruledtabular}
\end{table}

We have also studied the mean field theory based on the Lagrangian
Eq.~(\ref{U}) for comparison. It gives rise to a dual of mean
field theories with particle number constraints $\langle \hat
N_i\rangle=1$ and $\langle \hat N_i\rangle=2$, respectively. There
is a one-to-one correspondence between the solutions of the two
mean field theories and we will only consider the ``particle"
representation $\langle \hat N_i\rangle=1$. The mean field
equations are the same as Eq.~(\ref{MFEQ}) except that
Eq.~(\ref{MFEQc}) becomes
\[\langle \hat N_i\rangle=\frac{3}{2N}(1-\sum_k \frac{\chi_k}{E_k})=1. \]
The solution is also summarized in Table \ref{tab:spin1}. The
excitation spectrum is gapped but no longer dispersionless. We
note that the mean field solution Eq.~(\ref{MFEQ}) with
particle-hole symmetry has a better ground state energy.

\subsection{half-odd-integer spin: $S=3/2$}

For the spin-3/2 Heisenberg chain, we consider the effective
Hamiltonian Eq.~(\ref{Heff2}) with $b({3\over2})=0$ such that the
grounds state is a singlet state. We shall also take
$a({3\over2})={15\over16}$ for reason we shall see later. In
momentum space, the effective Hamiltonian becomes
\begin{eqnarray*}\label{H3/2k}
H_m&=&\sum_{ k} \chi_k
(c^\dagger_{\frac{3}{2}k}c_{\frac{3}{2}k}+c^\dagger_{\frac{1}{2}k}c_{\frac{1}{2}k}
+c^\dagger_{-\frac{1}{2}k}c_{-\frac{1}{2}k}+c^\dagger_{-\frac{3}{2}k}c_{-\frac{3}{2}k})\nonumber\\
&&+\sum_{ k} [\Delta_k^*(c_{-\frac{3}{2}-k}c_{\frac{3}{2}k} -
c_{-\frac{1}{2}-k}c_{\frac{1}{2}k})+h.c.]+\text{const.},
\end{eqnarray*}
where $\chi_k=\tilde\lambda_z-\frac{15}{8}J\chi\cos k$ and
$\Delta_k^*=\tilde\lambda_x-i\tilde\lambda_y-
\frac{15}{8}J\Delta^*\cos k$. We shall set $\chi$ and $\Delta$ to
be real numbers in the following.

As in the $S=1$ case, the Hamiltonian can be diagonalized by
Bogoliubov transformations and lead to the self consistent mean
field equations at zero temperature,
\begin{eqnarray}\label{MF3/2}
&&\chi=\frac{2S+1}{2N}\sum_k\cos k(1-\frac{\chi_k}{E_k}),\nonumber\\
&&\Delta=\frac{2S+1}{2N}\sum_k\cos k\frac{\Delta_k}{E_k},\nonumber\\
&&\langle\hat{M_+}\rangle=\frac{2S+1}{N(2S-1)}\sum_k
\frac{\Delta_k}{E_k}=\frac{(1-\tilde\lambda\coth\tilde\lambda)\tilde\lambda_+} {\tilde\lambda^2},\nonumber\\
&&\langle\hat{M}_z\rangle=\frac{2S+1}{N(2S-1)}\sum_k
\frac{\chi_k}{E_k}
=\frac{(1-\tilde\lambda\coth\tilde\lambda)\tilde\lambda_z}{\tilde\lambda^2},
\end{eqnarray}
where $S=3/2$, $\hat M_+=\hat M_x+i\hat M_y$,
$\tilde\lambda_+=\tilde\lambda_x+ i\tilde\lambda_y$ and
\begin{eqnarray}\label{Ex3/2}
E_k=\sqrt{(\tilde\lambda_z-{15\over8}J\chi\cos k)^2+|\tilde\lambda_+-{15\over8}J\Delta\cos k|^2}
\end{eqnarray}
is the spinon dispersion.

Solving the equations we find that $\pmb{\tilde\lambda}=0$ (i.e.
the ground state does not break particle-hole symmetry) and there
exists infinite degenerate solutions for $\chi$ and $\Delta$
satisfying $\sqrt{\chi^2+|\Delta|^2} =1.2732$ (see Table II). This
is a direct consequence of the $SU(2)$ gauge symmetry for half-odd
integer spins we mentioned in Section III.B.2, and all these
solutions are equivalent. Notice that the ground state energy
$E_0= -\frac{15NJ}{16} (\chi^2+|\Delta|^2)$ is the same as the
expectation value of the Heisenberg Hamiltoian Eq.~(\ref{H3/2}),
which is why we should choose $a({3\over2})=\frac{15}{16}$ in the
effective Hamiltonian. As a result of the particle-hole
symmetry($\pmb{\tilde\lambda}=0$), the spinon energy dispersion
Eq.~(\ref{Ex3/2}) is gapless at Fermi points $k=\pm\pi/2$.  The
spin-spin correlation is given by
\begin{eqnarray*}
\langle\mathbf S_i\cdot\mathbf S_{i+r}\rangle
&=&\frac{15}{4N^2}\sum_{p,q}e^{i(q-p)r}\left[(1+\frac{\chi_p}{E_p})
(1-\frac{\chi_q}{E_q})\right.\\&&\left.-\frac{\Delta_p^*}{E_p}
\frac{\Delta_q}{E_q}\right].
\end{eqnarray*}
  and decays at large distance as
$\langle\mathbf S_i\cdot\mathbf S_{i+r}\rangle\propto r^{-2}$
because of linearized spectrum around Fermi surface. This is also
confirmed directly by numerical calculation.

Similar to the $S=1$ case, we have also solved the mean field theory
with particle number constraint $\langle\hat N_i\rangle=1$ or
equivalently $\langle\hat{\mathbf M}\rangle=(0,0,1)^T$. The solution
is listed in Table \ref{tab:spin3/2}. Since $\tilde\lambda_z\neq0$
in this case, the spinon dispersion
$E_k=\frac{15J}{8}\sqrt{(\lambda_z/2-\chi\cos k)^2+|\Delta\cos
k|^2}$ breaks particle-hole symmetry and has a finite gap over the
whole Brillouin zone. The particle-hole symmetric solution is also
found to has a lower mean field ground state energy.

\begin{table}
\caption{Solutions of two versions of mean fields for spin-3/2
Heisenberg chain. In the first row ($\chi,\Delta$) means the real
combinations satisfying
$\sqrt{\chi^2+|\Delta|^2}=1.2732$}\label{tab:spin3/2}
\begin{ruledtabular}
\begin{tabular}{ccccccc}
$\langle N_i\rangle$      &    $\chi$     & $\Delta$      & $\tilde\lambda_z$ &   $E(J)$       & $E_{gap}(J)$  &$\langle\mathbf S_i\cdot\mathbf S_i\rangle$\\
\hline           2        &     $\chi$    & $\Delta$      &  0          &$-{1.5198}$     & ${0}$         &     15/4                                  \\
                 1        &$0 $           &${1.1490}$     &   0.7181    &$-{1.2377}$     & ${0.6732}$    &     45/16                                 \\
\end{tabular}
\end{ruledtabular}
\end{table}

\subsection{Haldane's conjecture and Edge states}

Comparing the mean field energy dispersion Eq.~(\ref{Ex1}) and
Eq.~(\ref{Ex3/2}), we find that the excitation spectrum of spin-1
Heisenberg model is gapped whereas the excitation spectrum for the
spin-3/2 Heisenberg model is gapless in the mean field formulation
with particle-hole symmetry. This difference between integer spin
and half-odd-integer spin persists for any spin $S$, if we
consider the trial Hamiltonian Eq.~(\ref{Heff1}) or
Eq.~(\ref{Heff2}) with $a(S)\neq0, b(S)=0$, i.e. if we consider
BCS spin-singlet ground state wavefunctions. The mean field
equations for arbitrary $S$ are the same as Eq.~(\ref{MFEQ}) or
Eq.~(\ref{MF3/2}) for arbitrary $S$, and the dispersion is
qualitatively the same as Eq.~(\ref{Ex1}) or Eq.~(\ref{Ex3/2}), as
long as we consider particle-hole symmetric mean field solutions
which do not break translational invariance. The main effect of
changing $S$ in mean field theory is to change the number of
fermionic spinon species. This result is consistent with the
Haldane conjecture, \cite{Haldane} which asserts that the integer
spin Heisenberg chains have singlet ground state with finite
excitation gaps and exponentially decaying spin-spin correlation
functions, while half-odd-integer spin chains have singlet ground
states with gapless excitation spectrums and power law decaying
correlations. Our particle-hole symmetric mean field results agree
well with Haldane conjecture.

 Haldane noticed that the time reversal operator satisfies
$T^2=1$ for integer spin and $T^2=-1$ for half-odd-integer spin,
and this results in different Berry phase contributions from the
topological excitations (skymion or instanton) in the path
integral formulation given by $e^{i2\pi SQ}$, where
\begin{eqnarray}\label{tplg}
Q=\int dtdx{1\over4\pi}\pmb n\cdot(\partial_t\pmb
n\times\partial_x\pmb n)
\end{eqnarray}
is the Skyrmion number. For $Q=$ odd integer, the Berry phase
$(e^{i2\pi S})^Q$ is $1^Q=1$ for integer spins and $(-1)^Q=-1$ for
half-odd-integer spins. The difference in the $\pm1$ factor
between integer spin and half-odd-integer spin is the origin of
Haldane's conjecture. The situation is quite similar in our mean
field theory. Noticing that $\Delta_{ij}=\bar C_i^ \dagger C_j$ is
a singlet formed by two spin-$S$ particles. The Clebsch-Gordan
coefficients implies that $\Delta_{ji}=-\Delta_{ij}$ for integer
$S$ and $\Delta_{ji}=\Delta_{ij}$ for half-odd-integer $S$ (see
Appendix B). This sign or parity difference results in appearance
of $\Delta\sin k$ in Eq.~(\ref{Ex1}) for $S=1$ and $\Delta\cos k$
in Eq.~(\ref{Ex3/2}) for $S=3/2$, which leads to different
symmetries of ground state wavefunctions and different excitation
spectrums between integer and half-odd-integer spin chains.

\begin{figure}[htbp]
\centering \includegraphics[width=3.3in]{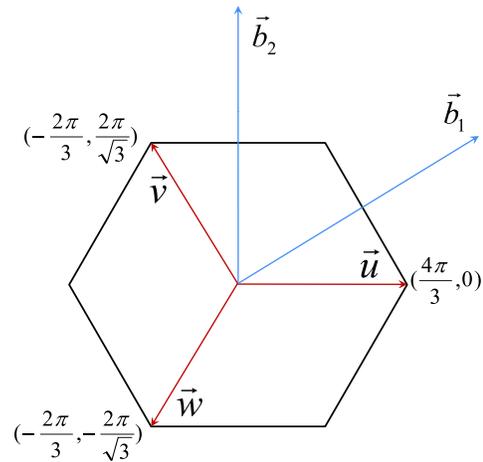} \caption{(Color
online) The first Brillouin zone of the triangular lattice. $\pmb
b_1=(2\pi, {2\pi\over\sqrt3})$ and $\pmb b_2=(0, {4\pi\over\sqrt3})$
are two reciprocal vectors. The Brillouin zone is divided into three
regular pieces by the three new bases $\pmb u$, $\pmb v$ and $\pmb
w$. }\label{fig:BZ}
\end{figure}

The topological term Eq.~(\ref{tplg}) leads to exponentially
localized edge states for open integer spin chains which is absent
for half-odd-integer spin chains. \cite{Ng} This important
difference between integer spin and half-odd-integer spin chains
is also reflected in our mean field theory where zero energy
Majorana edge fermions exist at the open boundaries for integer
spin chains which are absent for half-odd-integer spin chains. The
existence of Majorana edge fermions for integer spin chains is a
direct consequence of the odd-pairing symmetry for integer spin
chains. \cite{Read} The details of the Majorana edge fermions is
discussed in Appendix C.

\section{2D: $S=1$ spin liquids on triangular lattice}

We show in the previous section that our mean field theory is able
to capture the essential physics of spin-liquid states in 1D
antiferromagnetic quantum spin chains. In this section we shall
apply our mean field theory to the 2D $J_1$-$J_3$ Heisenberg model
on triangular lattice. We shall show that our mean field theory
admits new spin liquid solutions not explored before. Some results
of this model have been reported in a previous paper\cite{Liu},
where particle number constraint is treated in the ``particle
representation" ($\langle \hat N_i\rangle=1$). In this paper, we
shall revisit this model with the particle-hole symmetric
constraint. The Hamiltonian of the $J_1$-$J_3$ model is
\begin{eqnarray}\label{Hamiltonian}
H=\sum_{\langle i,j\rangle}J_1\mathbf  S_i\cdot\mathbf S_j +
J_3\sum_{[i,j]}\mathbf S_i\cdot \mathbf S_{j},
\end{eqnarray}
where $\langle i,j\rangle$ denotes nearest neighbor (NN) and
$[i,j]$ the next next nearest neighbors (NNNN). For $S=1$, there
are two channels of decoupling the spin interaction, namely,
$\mathbf S_i\cdot\mathbf
S_j=-\frac{J}{2}:\mathrm{Tr}(\psi_i^\dagger\psi_j
\psi_j^\dagger\psi_i):$ or $ \mathbf S_i\cdot\mathbf S_j=
-\frac{J}{2} :\mathrm{Tr}(\psi_i^\dagger\mathbf{I}\psi_j\cdot
\psi_j^\dagger \mathbf{I}\psi_i):$. Since there is no
Mermin-Wigner theorem to protect us at zero temperature at 2D, we
have to keep both terms in the effective Hamiltonian
Eq.~(\ref{Heff1}). We shall choose $b=1-a$, with the weight $a$
determined by minimizing the ground state energy.

\begin{figure}[htbp]
\centering \includegraphics[width=3.0in]{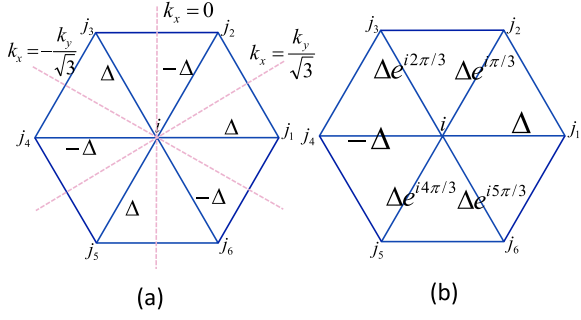}
\caption{(Color online) Two different pairing symmetries for the
singlet pairing $\Delta_{ij}$. We indicate the relative phase of
the nearest neighbor pairing $\langle\Delta_{ij}\rangle =\Delta_1
e^{i\varphi_{ij}}$. (a), $f$-wave pairing. The dished lines are
the zero lines of the pairing function (in lattice space or
momentum space). (b), $p_x+ip_y$-wave pairing. There are no zero
lines, the pairing function vanishes only at the $\Gamma$
point.}\label{fig:pairing}
\end{figure}

Next we introduce the mean field parameters $\chi_1$, $\Delta_1$,
$\mathbf u_1$, $\mathbf v_1$(for NN) and $\chi_3$, $\Delta_3$,
$\mathbf u_3$, $\mathbf v_3$ (for NNNN) to decouple the
Hamiltonian. We assume that $\mathbf u$ and $\mathbf v$ are
parallel, so we only keep the $z$ component of these vectors, and
note them simply as $u_{1,3}$ and $v_{1,3}$ (which are set as real
numbers). As in 1-D case we consider here only spin liquid
solutions that respect translation and rotational symmetries.
Since the pairings have odd parity $\langle\Delta_{ij}\rangle=
-\langle \Delta_{ji}\rangle$, only $p$-wave and $f$-wave like
pairings are allowed. Solving the mean field equations we find two
kinds of solutions (see Fig.~\ref{fig:pairing}), which have
$f$-wave and $p_x+ip_y$-wave pairing symmetries, respectively. We
shall concentrate on these two kinds of solutions in the
following.

In momentum space (Fig.~\ref{fig:BZ}), the mean field Hamiltonian
Eq.~(\ref{Heff1}) can be written as
\begin{eqnarray}\label{MF_hb}
H_m&=&\sum_{\mathbf{k}} \chi_k
(c^\dagger_{1k}c_{1k}+c^\dagger_{0k}c_{0k}+c^\dagger_{-1k}c_{-1k})\nonumber\\
&&-\sum_{\mathbf{k}} [\Delta_k^*(c_{-1-k}c_{1k}-\frac{1}{2}
c_{0-k}c_{0k})+h.c.]\nonumber\\&&+\sum_{\mathbf{k}} v_k
(c^\dagger_{1k}c_{1k}-c^\dagger_{-1k}c_{-1k})
\nonumber\\&&-\sum_{\mathbf{k}} u_{k}^*(c_{-1-k}c_{1k}+h.c.),
\end{eqnarray}
with
\begin{eqnarray*}
\chi_k&=&\tilde\lambda-a Z(J_1\chi_1\gamma_k+J_3\chi_3\gamma_{2k}),\nonumber\\
\Delta_k^*&=&ia Z(J_1\Delta_1\psi_k+J_3\Delta_3\psi_{2k}),\nonumber\\
v_k&=&-(1-a)Z(J_1\gamma_k v_1+J_3\gamma_{2k}v_3),\nonumber\\
u_k^*&=&(1-a)Z(J_1\gamma_k u_1+J_3\gamma_{2k} u_3),
\end{eqnarray*}
where $\tilde\lambda$ is the Lagrange multiplier and
\begin{eqnarray*}
&\gamma_k={1\over3}[\cos k_x+\cos(-{k_x\over2}+{\sqrt3
k_y\over2})+\cos(-{k_x\over2}-{\sqrt3k_y\over2})],\\
&\psi_k^{f}={1\over3}[\sin k_x+\sin(-{k_x\over2}+{\sqrt3
k_y\over2})
+\sin(-{k_x\over2}-{\sqrt3k_y\over2})],\\
&\psi_k^{p+ip}={1\over3}[\sin
k_x+e^{i{\pi\over3}}\sin({k_x\over2}+ {\sqrt3 k_y\over2})\ \ \ \ \
\ \ \ \ \ \  \ \ \ \ \  \ \ \ \ \ \
\\& \ \ \ \ +e^{i{2\pi\over3}} \sin(-{k_x\over2}+{\sqrt3
k_y\over2})].\ \ \ \ \ \ \ \ \ \ \  \ \ \ \ \  \ \ \ \ \ \
\end{eqnarray*}
The pairing symmetries can be verified easily by expanding the
pairing terms at small $k$: $\Delta^f_k \propto
k_x(k_x^2-3k_y^2)$, and $\Delta^{p_x+ip_y}_k\propto k_x+ik_y$.
There are three lines of zeros for the $f$-wave pairing function
as shown in Fig.~\ref{fig:pairing}.

Introducing a vector $C_k=(c_{1k},c^\dagger_{-1-k}, c_{0k},
c^\dagger_{0-k})^T$, the Hamiltonian Eq.~(\ref{MF_hb}) can be
written in a matrix form:
\begin{eqnarray}
H_m=\sum_{\mathbf{k}}C_k^\dagger H_k
C_k+\sum_{\mathbf{k}}(\frac{3}{2}\chi_{k}-v_{k}),
\end{eqnarray}
where
\begin{eqnarray}
H_k=\left(\begin{matrix} \chi_k+v_k  & -u^{}_k-\Delta_k & 0
&0\\
-u^{*}_k-\Delta_k^*  & -\chi_{k}+v^{}_{k} & 0 &0\\ 0 &0 & {\chi_k\over2} & {\Delta_k\over2}\\
0  & 0 & {\Delta_k^*\over2} & -{\chi_{k}\over2}
\end{matrix}\right).\nonumber
\end{eqnarray}
$H_k$ can be diagonalized by the Bogoliubov transformation,
\begin{eqnarray}\label{Boglbv}
&&A_{k}=\cos\frac{\theta_k}{2}c_{1k} -\sin\frac{\theta_k}{2}e^{i\eta_k} c_{-1-k}^\dagger,\nonumber\\
&&B_{-k}^\dagger=\sin\frac{\theta_k}{2}e^{-i\eta_k}c_{1k} +\cos\frac{\theta_k}{2} c_{-1-k}^\dagger,\nonumber\\
&&D_k=\cos\frac{\Theta_k}{2}c_{0k} + \sin\frac{\Theta_k}{2}e^{i\phi_k} c_{0-k}^\dagger,
\end{eqnarray}
where $\tan\theta_k={|u_k+\Delta_k|\over\chi_k},
e^{i\eta_k}={u_k+\Delta_k\over|u_k+\Delta_k|},
\tan\Theta_k={|\Delta_k|\over\chi_k}$ and
$e^{i\phi_k}={\Delta_k\over|\Delta_k|}$. The corresponding
eigenvalues are given by $v_k\pm E_{1k}$ and $\pm E_{0k}$, where
\[ E_{1k}=\sqrt {\chi_k^2+|\Delta_k|^2+u_k^2}, E_{0k}=\sqrt{\chi_k^2+|\Delta_k|^2},  \]
and the self-consistent mean field equations are
\begin{eqnarray}\label{MFeqs1}
&&\bar
\chi_{1,3}=\frac{1}{N}\sum_{\mathbf{k}}\gamma_{k,2k}[\frac{3}{2}-\frac
{\chi_k}{E_{1k}}- \frac{\chi_k}{2E_{0k}}], \nonumber\\
&&\bar \Delta_{1,3}=\frac{1}{N}\sum_{\mathbf{k}}\psi_{k,2k}[\frac
{|\Delta_k|}{E_{1k}}+ \frac{|\Delta_k|}{2E_{0k}}], \nonumber\\
&&\bar u_{1,3}=\frac{1}{N}\sum_{\mathbf{k}}\gamma_{k,2k}\frac
{u_k}{E_{1k}}, \nonumber\\
&&\bar v_{1,3}=0, \nonumber\\
&&\langle \hat N_i-\frac{3}{2}\rangle=\frac{1}{2}\tanh
\frac{\tilde\lambda}{2}, 
\end{eqnarray}
where we have adopted the particle-hole symmetric constraint in
writing down the mean field equations. We have also checked the
mean field solutions with constraint $\langle \hat
N_i\rangle=1$\cite{Liu} and found that the excitation spectrums
are qualitatively the same as those in the particle-hole symmetric
theory. The ground state energy is lowest for vanishing $u_{1,3},
v_{1,3}$, i.e. $a=1$, $u_k=v_k=0$ and the ground state is a
spin-singlet. The excitations are characterized by three branches
of fermionic spinons with $S_z=0,\pm1$ and identical dispersion
$E_k=\sqrt{\chi_k^2 +|\Delta_k|^2}$.

\begin{figure}[htpb]
\centering \includegraphics[width=2.8in]{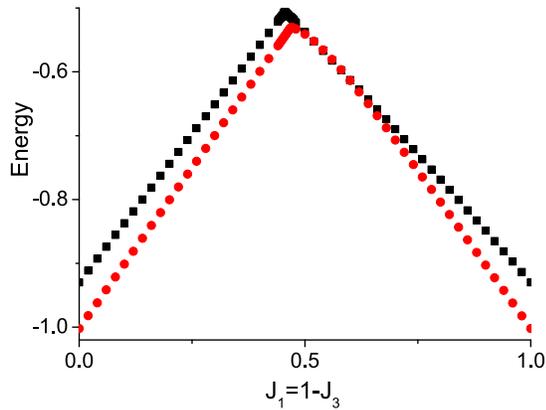} \caption{(Color
online)Phase diagram for the $J_1$-$J_3$ Heisenberg model with
$J_1+J_3=1$. We use the constraint $\langle \hat
N_i-\frac{3}{2}\rangle=\frac{1}{2}\tanh \frac{\tilde\lambda}{2}$.
The red dot line indicate energy per site for the $p_x+ip_y$ state
and the black square line is for the f-wave state. Dirac nodes
exist for the f-wave state. The number of nodes is 6 when $J_1$ is
dominating and 24 when $J_3$ is dominating.}\label{fig:HSB}
\end{figure}

Our mean field theory for the $J_1$-$J_3$ model contains two
regimes of spin liquid states for both $f$ and $p_x+ip_y$ pairing
symmetries as a function of $J_1/J_3$. A first order phase
transition occurs between the two regimes at $J_1/J_3\sim1$ (see
Fig.~\ref{fig:HSB}). When $J_1$ dominates, the spin liquid state
is characterized by $\chi_{1,3}\neq0$ and $\Delta_{1,3}\neq0$
(consequently $\langle\mathbf S_i\cdot\mathbf S_{i+1}\rangle<0$
and $\langle\mathbf S_i\cdot\mathbf S_{i+2}\rangle<0$); while when
$J_3$ dominates, $\chi_1=\Delta_1=0$ and $\chi_3\neq0$,
$\Delta_3\neq0$ (consequently $\langle\mathbf S_i\cdot\mathbf
S_{i+1}\rangle=0$, $\langle\mathbf S_i\cdot\mathbf
S_{i+2}\rangle<0$). The $p_x+ip_y$ states remain lower in energy
in both regimes.

\begin{figure}[htbp]
\centering \includegraphics[width=3.3in]{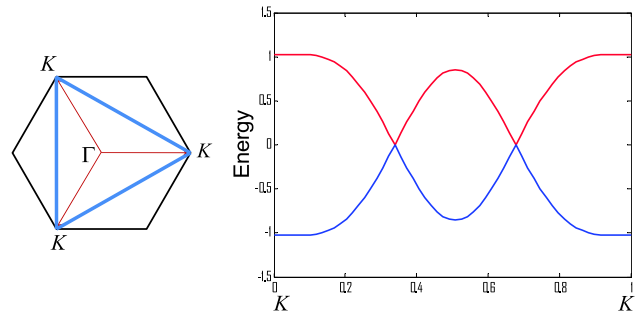}
\caption{(Color online) The spinon dispersion the f-wave pairing
state with constraint $\langle \hat
N_i-\frac{3}{2}\rangle=\frac{1}{2}\tanh \frac{\tilde\lambda}{2}$
for $J_1=0.6$, $J_3=0.4$. The figure on the right shows the
dispersion along the line linking two $K$ points of the Brillouin
zone (see the blue lines on the left). The negative energy part is
filled and the excitation is gapless with 6 Dirac nodes in the
first Brillouin zone. The dispersion for the mean field solution
with $\langle\hat N_i\rangle=1$ is similar to the above one except
that the locations of the nodes are shifted a little bit.}
\label{fig:f_band}
\end{figure}

The $f$-wave pairing solution respects particle-hole symmetry and
has $\langle\hat N_i\rangle={3\over2}$. The excitation is gapless
with several Dirac cones in the first Brillouin zone. A cut of the
spinon dispersion is shown in Fig.~\ref{fig:f_band} where the
particle-hole symmetry is obvious. The mean field solution with
$\langle\hat N_i\rangle=1$ has similar properties, except that the
particle-hole symmetry is lost and the position of the Dirac nodes
are shifted.\cite{Liu}

For the $p_x+ip_y$-wave pairing, we find that the ground state
breaks particle-hole symmetry and $\tilde\lambda\neq0$. The
excitation spectrum is fully gapped. Similar to integer spin
chains, we find that the $p_x+ip_y$-wave state is topologically
nontrivial (see also Appendix C). The solution has
$|\tilde\lambda| < Z|J_1\chi_1+J_3\chi_3|$, so that $\chi_k$ can
be either positive or negative, depending on $\mathbf{k}$.
Therefore the Bogoliubov spinor space described by the vector
$({Im\Delta_k\over E_k}, {Re\Delta_k\over E_k},{\chi_k\over
E_k})^T$ (which is $S^2$) covers the Brillouin zone ($\mathbf{k}$
space, which is also $S^2$) at least once. In other words, the
topological (Skyrmion) number of mapping from the $\mathbf{k}$
space to the spinor space is nonzero $m\neq0$. However in the
vacuum where the spinon density is zero, $\tilde \lambda$ is very
big and $\chi_k$ can only take positive values, which gives a zero
topological number. \cite{Read} Since the bulk and vacuum belong
to different topological sectors, the boundary defines a domain
wall between the two phases, and there should exists gapless
(chiral) Majorana edge states in the $p_x+ip_y$ state following
the analysis of Read and Green. \cite{Read} In this sense, the
$p_x+ip_y$-state describes a time-reversal symmetry breaking
topological spin liquid.

Next we consider the specific heat and spin susceptibility for the
$f$- and $p_x+ip_y$ states. For the $p_x+ip_y$-wave pairing
states, since the excitations are fully gapped as in BCS
superconductors, the specific heat and spin susceptibility will
show exponential behavior \cite{Tinkham} at low temperature. For
the $f$-wave pairing states, both the specific heat and the spin
susceptibility show power law behavior due to the Dirac cone
structure (See Fig.~\ref{fig:f_band}). The energy for the system
is $E=3\sum_k\frac{E_k} {e^{\beta E_k}+1}$ in mean field theory.
At low temperature, the specific heat is dominate by the low
energy excitations near the nodes of the fermi surface, which is
given approximately by $E_k=v_Fk$, where $v_F$ is the fermi
velocity (the anisotropy of the dispersion can be removed by a
re-scaling of momentum and will not affect the result
qualitatively). The magnetic specific heat is given by
\begin{eqnarray}
C_M=\frac{\partial E}{\partial T}&=&3\frac{\partial}{\partial
T}\int_0^{k_\Lambda}\frac{v_Fk}{e^{\frac{v_Fk}{T}}+1} 2\pi kdk\nonumber\\
&=&3\frac{\partial}{\partial
T}\int_0^{\frac{v_Fk_\Lambda}{T}\to\infty}
\frac{1}{e^{x}+1} 2\pi x^2\frac{T^3}{v_F^2}dx\nonumber\\
&\propto&n\pi T^2/v_F^2
\end{eqnarray}
where $n$ is the number of Dirac cones.  The magnetic
susceptibility can be calculated from the linear response theory
with
\begin{eqnarray}\label{kubo}
\chi_z(k,i\omega)&=&\int d^2r\int d\tau T_\tau\langle
S_z(r,\tau)S_z(0,0)\rangle e^{i(\omega\tau-kr)}\nonumber\\
&=&\frac{2}{N^2}\int d^2r\int d\tau e^{i(\omega\tau-kr)}
\sum_{q,q',p,p'}e^{i(q'-q)r}\nonumber\\
&&\times[T_\tau\langle c_1^\dagger(q,\tau)c_1(q',\tau)
c_1^\dagger(p,0)c_1(p',0)\rangle \nonumber\\&&-T_\tau\langle
c_1^\dagger(q,\tau)c_1(q',\tau) c_{-1}^\dagger
(p,0)c_{-1}(p',0)\rangle ]\nonumber\\
\end{eqnarray}
The static susceptibility is given by
$\chi_z=\lim_{k\rightarrow0}\chi_z(k,0) \propto {T\over v_f^2}$.
The detailed calculation is given in Appendix D.

Recently, a spin-1 material $\mathrm {NiGa_2S_4}$ was discovered,
where the $S=1$ $\mathrm{Ni}^{2+}$ ions form a triangular lattice
with antiferromagnetic (AFM) interaction. The system has a
Curie-Weiss temperature $\sim80$K, and is found to exhibit no
conventional magnetic order down to $0.35$K. Several plausible
ground states have been proposed for this system,
\cite{science,PRBexp,AFQ,SU(3),Qphase,J1J3K,Vorex} including an
antiferro-nematic (AFN) state \cite{AFQ}, a ferro-nematic state,
\cite{PRBexp} and a KT phase driven by vortices. \cite{Vorex} Spin
liquid is also proposed to be a plausible ground state. In our
mean field theory, the $f$-wave pairing spin liquid has coherent
gapless excitation with specific heat scaling as $T^2$ at low
temperature which is consistent with the experimental result.
\cite{science} Furthermore, the Knight shift data in
Ref.~\onlinecite{PRBexp} shows that the intrinsic susceptibility
decreases with decreasing temperature (but the very low
temperature data are missing. \cite{PRBexp}) This matches
qualitatively with our mean field result Eq.~(\ref{kubo}). We
therefore proposed that the $f$-wave pairing spin liquid state is
a plausible ground state for the material.

\section{Comments and Conclusion}

 Summarizing, we show in this paper how the fermionic
representation for spin $S=1/2$ systems can be generalized to spin
systems with $S>1/2$. The symmetry group of the spin operator is
$SU(2)$ group for integer spin and $U(1)\bar\otimes Z_2$ group for
half-odd-integer spin. Different path integral formulations and
mean field theories are developed corresponding to different ways
of handling the constraints. In 1D, we show that the particle-hole
symmetric mean field theory for $S=1$ and $S=3/2$ spin chains are
consistent with Haldane's conjecture, and we argue that the
difference reflects a fundamental difference between integer and
half-odd-integer spin chains. We also study 2D spin-1 AFM on
triangular lattice where we find two spin liquid states, one is a
gapless $f$-wave spin liquid and the other is a topological
$p_x+ip_y$ spin liquid state. We propose that the gapless $f$-wave
spin liquid is a plausible ground state for the material
NiGa$_2$S$_4$. Our approach can be applied to any other spin
models and provides a new approach to spin liquid states for
$S>1/2$ spin systems.

We note that because of limitation in scope we have addressed only a
very limited number of issues in the study of spin-liquid states in
this paper. Within the mean-field theory we have restricted
ourselves to spin liquids solutions of Hamiltonian
(\ref{Hamiltonian}) with spin rotational symmetry (1D) and lattice
translational symmetry (1D and 2D). We note that solutions which
break spin rotational symmetry exist in our theory which may serve
as ground states of Hamiltonian (\ref{Hamiltonian})\cite{Liu}. For
example, we show in Ref.~\onlinecite{Liu} that state with
long-ranged magnetic order exists as lowest energy state of the
mean-field theory of the $J_1$-$J_3$ model and anti-ferro nematic
order may exist when the bi-quadratic ($K(\mathbf{S}_i\cdot
\mathbf{S}_j)^2$) spin-spin interaction exists in the Hamiltonian.
For simplicity we only considered the mean field states without
breaking the translational symmetry. We note that states that break
translational symmetry (such as dimerized states) are believed to be
ground states of some 1D or 2D spin models.\cite{dimer}

Another very important issue is whether the spin liquid states we
find are stable against gauge fluctuations. In 1-D, the gauge
fluctuations can be removed by a time-dependent Read-Newns gauge
transformation and thus have no effect to the low energy
properties\cite{MF}. The situation is very different in two or
higher dimensions where the stability of the mean-field state
depends on dimensionality and the (gauge) structure of the gauge
field fluctuations. For instance, $Z_2$ spin liquid is believed to
be stable at 2D\cite{Z2} and gapless $U(1)$ Dirac fermionic spin
liquids is stable at 2D in the large-$N$ limit\cite{Hermele}. In our
case of spin liquid solutions for the $S=1$ Heisenberg model, the
$U(1)$ gauge fluctuation is gapped by the spinon-pairing term via
the Anderson-Higgs mechanism. The ground state is stable against the
low energy gauge fluctuations which is described by an effective
$Z_2$ gauge theory. We note that the emergence of gauge-field
structure is a direct consequences of particle number constraints
and a reliable answer to the question of stability of spin liquid
states can be obtained only if we can handle the particle number
constraint reliably. Gauge field theory can only handle long
distance, low energy gauge fluctuations and a more satisfactory
answer to the question of stability of the spin liquid states can be
obtained only after the Gutzwiller Projection wavefunctions are
studied carefully.

We thank Prof. Michael Ma for the discussion about 1D models. We
also thank Prof. Patrick. A. Lee and Naoto Nagaosa for suggesting
the $p_x+ip_y$-ansatz for AFM on triangular lattice, and thank Mr.
Cheung Chan and Dr. Xiao-Yong Feng for helpful discussions. ZXL and
TKN are supported by RGC grant of HKSAR and HKUST3/CRF/09. YZ is
supported by National Basic Research Program of China (973 Program,
No.2011CB605903), the National Natural Science Foundation of
China(Grant No.11074218) and the Fundamental Research Funds for the
Central Universities in China.

\appendix
\section{Symmetry group of the spin operators and the pairing parity}

We shall look for allowed transformations $\psi\to\psi W $, which
keep the spin operator Eq.~(\ref{spin}) invariant. At first
glance, this condition is satisfied as long as $W\in U(2)$.
However, it is {\em not true} because $c_m$'s and $c_m^\dagger$'s
are not independent. By checking each row of $\psi$ directly, it
is easy to see that $(c_m,(-1)^{S-m}c^\dagger_{-m})$ transforms to
$(c_m,(-1)^{S-m}c^\dagger_{-m}) W$, and
$(c^\dagger_m,(-1)^{S-m}c_{-m})\rightarrow(c^\dagger_m,(-1)^{S-m}c_{-m})
W^*$. Noticing that
$(c_{m}^\dagger,(-1)^{S-m}c_{-m})=(-1)^{S-m}(c_{-m},(-1)^{S-m}c^\dagger_{m})\sigma_x$,
 we obtain
\begin{subequations}
\begin{eqnarray*}\label{W1}
(-1)^{S-m}(c_{-m},(-1)^{S-m}c^\dagger_{m})\sigma_x\to\nonumber\\
(-1)^{S-m}(c_{-m},(-1)^{S-m}c^\dagger_{m})\sigma_x W^*.
\end{eqnarray*}
or
\begin{eqnarray}
&(c_{-m},(-1)^{S-m}c^\dagger_{m})\to&\nonumber\\
&(c_{-m},(-1)^{S-m}c^\dagger_{m})\sigma_x
W^*\sigma_x.&
\end{eqnarray}

On the other hand, $(c_{-m},(-1)^{S+m}c^\dagger_{m})$ transforms
to $(c_{-m},(-1)^{S+m}c^\dagger_{m})W$. Noticing that
$(c_{-m},(-1)^{S+m}c^\dagger_{m})=(c_{-m},(-1)^{-S-m}c^\dagger_{m})
=(c_{-m},(-1)^{S-m}c^\dagger_{m})(\sigma_z)^{-2S}$ and
$(\sigma_z)^{-2S}=(\sigma_z)^{2S}$, we obtain
\begin{eqnarray*}
(c_{-m},(-1)^{S-m}c^\dagger_{m})(\sigma_z)^{2S}\to\nonumber\\
(c_{-m},(-1)^{S-m}c^\dagger_{m})(\sigma_z)^{2S}W.
\end{eqnarray*}
or
\begin{eqnarray}\label{W2}
&(c_{-m},(-1)^{S-m}c^\dagger_{m})\to&\nonumber\\
&(c_{-m},(-1)^{S-m}c^\dagger_{m})(\sigma_z)^{2S}W(\sigma_z)^{2S}.&
\end{eqnarray}
\end{subequations}
Eq.~(\ref{W1}) and Eq.~(\ref{W2}) impose the condition
\begin{equation}
\sigma_xW^*\sigma_x=(\sigma_z)^{2S}W(\sigma_z)^{2S}. \label{W3}
\end{equation}

For integer spin, Eq.~(\ref{W3}) becomes $W^*=\sigma_x W\sigma_x$,
which results in $W=e^{i\theta\sigma_z}$ or $W=e^{i\theta
\sigma_z} \sigma_x$, i.e. $W\in U(1)\bar\otimes Z_2$, where
$U(1)=\{e^{i\sigma_z\theta}\}$ is the usual gauge symmetry group
and $Z_2=\{\mathbb{I},\sigma_x\}$ is the particle-hole symmetry
group. Notice that the two groups don't commute, thus the total
symmetry group is a semidirect product of the $U(1)$ group and the
$Z_2$ group $U(1)\bar\otimes Z_2=\{e^{i\sigma_z\theta}, \sigma_x
e^{i\sigma_z\theta}=e^{-i \sigma_z\theta}\sigma_x;\theta\in
\mathbb{R}\}$.  For half-odd-integer spin, Eq.~(\ref{W3}) implies
$W^*=\sigma_y W\sigma_y$, which is equivalent to $\det W=1$ since
$W^\dagger W=1$. This indicate that $W\in SU(2)$. Thus the
symmetry group is $SU(2)$ group.

\section{proof of some algebraic relations}
First we introduce a $(2S+1)\times(2S+1)$ matrix $B$, which is
essential in our proof:
\begin{eqnarray}\label{B}
B=\left(\begin{matrix}&&&&1\\&&&-1&\\&&1&&\\\cdots&&&&\end{matrix}\right).
\end{eqnarray}
The matrix is the CG coefficients combining two spin-S to a
spin-singlet state, i.e. $|0,0\rangle=B_{mm'}|S,m\rangle_1
|S,m'\rangle_2=C^\dagger \bar C|vac\rangle$. Thus $\bar C$ can be
expressed as $\bar C=BC^*=B(c_{S}^\dagger, c_{S-1}^\dagger,
\cdots,c_{-S}^\dagger)^T$. Notice that for integer spin $B^{-1}=B$
while for half-odd-integer spin $B^{-1}=-B$.

The following are some common properties of the $B$ matrix.
Suppose $\hat R$ is an rotation operation, and $D(R)$ is its
$(2S+1)\times(2S+1)$ dimensional matrix representation, then we
have $\hat R C=D(R)C$, and
\[\hat R \bar C=D(R)\bar C=D(R)BC^*.\]
On the other hand, $\hat RC^*=D(R)^* C^*$, so
\[\hat R\bar C=\hat R(BC^*)=B\hat RC^*=BD(R)^* C^*.\]
Comparing these two, we have $D(R)B=BD(R)^*$, or
\begin{eqnarray}\label{BDB}
B^{-1}D(R)B=D(R)^*.
\end{eqnarray}
\textit{This means that $B$ is the transformation matrix which
transfers the representation matrix of a rotational operator to
its complex conjugate.} In general, $D(R)$ can be written as
$D(R)=e^{i\mathbf I\cdot\mathbf R}$, where $\mathbf R$ are the
$SU(2)$ group parameters (real numbers) for the operator $\hat R$,
and consequently $D(R)^*=e^{-i\mathbf I^*\cdot\mathbf R}$. From
Eq.~(\ref{BDB}) we have
\begin{eqnarray*}
B^{-1}D(R)B&=&B^{-1}e^{i\mathbf I\cdot\mathbf R}B=e^{iB^{-1}\mathbf IB\cdot\mathbf R}\\
&=&D(R)^*=e^{-i\mathbf I^*\cdot\mathbf R}.
\end{eqnarray*}
So we have \[B^{-1}\mathbf IB=-\mathbf I^*=-\mathbf I^T.\]

Noticing $B^{-1}=\pm B$, above equation is equivalent to $B\mathbf
IB^{-1}=-\mathbf I^*=-\mathbf I^T.$

Now let us consider the spin algebra. The spin interaction is
decomposed by the following two operators
\begin{eqnarray}
\psi_i^\dagger\psi_j&=&\left(\begin{matrix}C_i^\dagger C_j & C_i^\dagger \bar C_j\\
\bar C_i^\dagger C_j & \bar C_i^\dagger\bar C_j\end{matrix}\right),\label{CD}\\
\psi_i^\dagger\mathbf I\psi_j&=& \left(\begin{matrix}C_i^\dagger \mathbf IC_j & C_i^\dagger\mathbf I \bar C_j\\
\bar C_i^\dagger\mathbf I C_j & \bar C_i^\dagger\mathbf I\bar
C_j\end{matrix}\right).\label{uv}
\end{eqnarray}
We define $\chi_{ij}=C_i^\dagger C_j$, $\Delta_{ij}=\bar
C_i^\dagger C_j$ and $\mathbf u_{ij}=\bar C_i^\dagger\mathbf I
C_j$, $\mathbf v_{ij}=C_i^\dagger \mathbf IC_j$. The following
discussions distinguish integer- and half-odd-integer- spins and
prove the results in section III.C.

{\bf{integer spins}} ($B^\dagger=B^T=B^{-1}=B$)

The second term on the first row of Eq.~(\ref{CD}) can be written
as
\begin{eqnarray*}
C_i^\dagger\bar C_j&=&C_i^\dagger BC_j^*=-(C_i^\dagger
BC_j^*)^T\nonumber\\&=&-C_j^\dagger BC_i^*=-C_j^\dagger\bar
C_i=-\Delta_{ij}^\dagger
\end{eqnarray*}
In second step, we have used the fact that the transverse of a
scalar operator is equivalent to reversing the order of the
constituting components. Commuting two fermionic operators we
obtain a minus sign. On the other hand, $\Delta_{ji}^\dagger=(\bar
C_j^\dagger C_i)^\dagger=C_i^\dagger\bar C_j$, so
$\Delta_{ji}=-\Delta_{ij}$.

The second term on the second row of Eq.~(\ref{CD}) is
\begin{eqnarray*}
\bar C_i^\dagger\bar C_j=C_i^TBBC_j^*=-(C_i^TC_j^*)^T
=-C_j^\dagger C_i=-\chi_{ij}^\dagger
\end{eqnarray*}
It is obvious that $\chi_{ji}=\chi_{ij}^\dagger$.

To simplify Eq.~(\ref{uv}), we will use the following relations:
$B\mathbf IB=-\mathbf I^T$, $B\mathbf I=-\mathbf I^TB$ and
$\mathbf IB=-B\mathbf I^T$. The second term on the first row of
Eq.~(\ref{uv}) is
\begin{eqnarray}
C_i^\dagger\mathbf I \bar C_j&=&C_i^\dagger\mathbf
IBC_j^*=-C_i^\dagger B\mathbf I^TC_j=(C_i^\dagger B\mathbf
I^TC^*_j)^T \nonumber\\  &=&C_j^\dagger\mathbf
IBC_i^*=C_j^\dagger\mathbf I\bar C_i=(\bar C_i^\dagger \mathbf
IC_j)^\dagger=\mathbf u_{ij}^\dagger\label{uij}
\end{eqnarray}
On the other hand, above operator can be written into another
form, $C_i^\dagger\mathbf I\bar C_j=(\bar C_j^\dagger\mathbf
IC_i)^\dagger=\mathbf u_{ji}^\dagger$. Comparing with
Eq.~(\ref{uij}) we get $\mathbf u_{ij}=\mathbf u_{ji}$.

The second term on the second row of Eq.~(\ref{uv}) is
\begin{eqnarray}
\bar C_i^\dagger\mathbf I\bar C_j&=&C_i^TB\mathbf
IBC_j^*=-C_i^T\mathbf I^TC_j^*\nonumber\\
&=&(C_i^T\mathbf I^TC^*_j)^T=C_j^\dagger\mathbf IC_i=\mathbf
v_{ij}^\dagger
\end{eqnarray}
And the relation $\mathbf v_{ji}^\dagger=\mathbf v_{ji}$ manifests
itself.

{\bf{half-odd-integer spins}} ($B^\dagger=B^T=B^{-1}=-B$)

Repeating the above procedures, it is straightforward to show that
\begin{eqnarray}
C_i^\dagger\bar C_j&=&\Delta_{ij}^\dagger,  \bar C_i^\dagger\bar
C_j=-\chi_{ij}^\dagger,\nonumber\\
C_i^\dagger\mathbf I \bar C_j&=&-\mathbf u_{ij}^\dagger,   \bar
C_i^\dagger\mathbf I\bar C_j=\mathbf v_{ij}^\dagger,
\end{eqnarray}
with $ \chi_{ji}=\chi_{ij}^\dagger,\ \Delta_{ji}=\Delta_{ij},
\mathbf u_{ji}=-\mathbf  u_{ij},\ \mathbf  v_{ji}=\mathbf
v_{ij}^\dagger$.

\section{Edge states in open Integer spin chains}

It is known that integer spin antiferromagnetic Heisenberg model
has free edge states with spin magnitude-S/2. \cite{Ng} Majorana
fermion edge states also exist in our mean field theory under open
boundary condition. Our discussion will follow the argument of
Ref.~\onlinecite{Read} for 2D pairing fermions. Firstly, we point
out the existence of two topologically distinct phases under
periodic boundary condition and then we discuss the zero energy
edge state solution at an open boundary.

In our mean field theory, all the properties of the ground state
are completely determined by $u_k$ and $v_k$ (or $\chi_k$ and
$\Delta_k$). Noticing that $u_k$ and $v_k$ obey
$|u_k|^2+|v_k|^2=1$, so they can be viewed as a spinor. This
spionor is equivalent to a vector
$(u^*,v^*)\pmb\sigma(u,v)^T=(0,\sin\theta_k, \cos\theta_k)^T=
(0,{\Delta_k^*\over iE_k},{\chi_k\over E_k})^T$ where we have used
Eq.~(\ref{ukvk}). Obviously, the spinor $(u,v)$ spans a $S^1$
space. Recalling that the first Brillouin zone is also $S^1$, the
functions $u_k, v_k$ describe a mapping from $S^1$(k space) to
$S^1$(spinor space). The mapping degree is characterized by the
first homotopy group $\pi_1(S^1)=\mathbf Z$. When $-2J\chi<\lambda
<2J\chi$ (which is the case for our mean field theory), the map is
nontrivial because every point of the spinor $S^1$ is covered
once, in other words, the mapping degree is $m=1$. This
topological number defines a phase, we call it A phase. When
$|\lambda|>2J\chi$, the mapping is topologically trivial since
$\chi_k$ is always positive(or negative) and the lower half
circle(or upper circle) of the spinor $S^1$ is never covered, so
the mapping degree is $m=0$. We call this region B phase. Since
the mapping cannot be smoothly deformed from $m=0$ to $m=1$, a
topological phase transition occurs at $|\lambda|=2J\chi$. We will
see that the existence of zero energy edge states is tied with a
phase boundary between the bulk (A phase) and the vacuum (B
phase).

Next we consider an infinite chain with a single edge. We assume
that the edge is located at $x=0$, and $x<0$ is the vacuum where
the spinon density is zero. The boundary can be described by a
potential $V(x)$ that is large and positive at $x<0$ so that the
wavefunction of the spinon vanishes exponentially at $x<0$.
Equivalently, we can assume a position dependent lagrange
multiplier term $\lambda(x)$ which becomes very large and
positive(so that the condition $|\lambda|>2J\chi$ is satisfied) at
$x<0$. This implies that the vacuum belongs to the B phase and the
boundary is a domain wall between A phase and B phase.

We now study the edge states in the continuum approximation. It is
convenient to introduce $\mu(x)=2J\chi-\lambda(x)$ such that $\mu$
is positive in the bulk, becomes zero near the edge and turns
negative outside the edge. Expanding $\chi_k$ and $\Delta_k$ to
first order in $k$ near $k=0$, we obtain $\chi_k\sim-\mu$ and
$\Delta_k\sim 2iJ\Delta k$. To see the effects of the domain wall,
we consider the BdG equation Eq.~(\ref{BdG-k}), which can be
written in position space as
\begin{eqnarray}\label{BdG}
i{\partial u\over\partial t} &=&-\mu
u-2J\Delta{\partial v\over\partial x},\nonumber\\
i{\partial v\over\partial t} &=&\mu v+2J\Delta{\partial
u\over\partial x},
\end{eqnarray}
where we have replaced $E$ and $k$ by $i{\partial\over\partial t}$
and $-i{\partial\over\partial x}$, respectively. The BdG equation
is compatible with $u(x,t)=v(x,t)^*$, the spinor $(u,v)$
satisfying this relation describes Majorana fermions. There is a
normalizeable zero energy solution for the above equations.
Putting $u=v$, then
\[2J\Delta{\partial u\over\partial x}=-\mu u,\]
and the solution is
\[u(x)=v(x)=\propto e^{-{1\over2J\Delta}\int^x \mu(x)dx}.\]
Notice that the BdG equation Eq.~(\ref{BdG})
\begin{eqnarray}
Eu&=&-\mu
u-2J\Delta{\partial v\over\partial x}\nonumber\\
Ev&=&\mu v+2J\Delta{\partial u\over\partial x}.\nonumber
\end{eqnarray}
at $E\neq0$ admits {\em no} normalizable bound solutions. This
shows that the zero mode is the only plausible bound state
solution.

In the case of a pair of boundaries at $x=0$ and $x=W$, we can
replace the two first order equations by a second order equation,
\[E^2(u\pm v)=[\mu^2-(2J\Delta)^2{\partial^2\over\partial x^2}
\mp(2J\Delta){\partial\mu\over\partial x}](u\pm v),\] and the
energy $E$ is not zero for finite $W$. However two bound state
solutions with energy going to zero exponentially with increasing
$W$ exist. The solution $u+v$ is centered at $x=0$ and $u-v$ is
centered at $x=W$. Notice that there is only one solution for the
above BdG equations, and the two modes $u\pm v$ are Majorana
fermions.

The above discussion is applicable to our mean field theory for
$S=1$ open antiferromagnetic chain. The discussion can be
generalized to larger integer spins $S>1$ if we adopt the
effective Hamiltonian Eq.~(\ref{Hm1}).

Unlike the $S=1$ case, our $S=3/2$ mean field theory produces a
mapping from $k$-space $S^1$ to the $(u,v)$ space $S^1$ which is
topologically trivial (essentially because of the even parity of
the pairing term), and the Majorana edge states don't exist
anymore. This is in agreement with the result that no
exponentially localized edge state exists for open
half-odd-integer spin chains. The existence of power-law localized
edge states in half-odd-integer spin chains \cite{Ng} is more
subtle and is not produced by our mean field theory.

\section{Calculation of the susceptibility}

Substituting the Bogoliubov transformation Eq.~(\ref{Boglbv}) into
the Kubo formula Eq.~(\ref{kubo}), we obtain
\begin{widetext}
\begin{eqnarray}
\chi_z(k,i\omega) &=&\frac{2}{V}\int d\tau e^{i\omega\tau}\sum_{q}
\left[[\mathcal C_q^2 G_A(q,-\tau)-\mathcal S_q^2 G_B(-q,\tau)]
[-\mathcal C_{q+k}^2 G_A(q+k,\tau)+\mathcal S_{q+k}^2
G_B(-q-k,-\tau)]\right.\nonumber\\&&\left. -\mathcal C_q\mathcal
S_q [G_A(q,-\tau) + G_B(-q,\tau)]\times\mathcal C_{q+k}\mathcal
S_{q+k}
[G_A(q+k,\tau)+ G_B(-q-k,-\tau)]\right]\nonumber\\
&=&\frac{2}{\beta V}\sum_{q,i\Omega} \left[[\mathcal C_q^2
G_A(q,i\Omega)-\mathcal S_q^2 G_B(-q,-i\Omega)] [-\mathcal
C_{q+k}^2 G_A(q+k,i\Omega+i\omega)+\mathcal S_{q+k}^2
G_B(-q-k,-i\Omega-i\omega)]\right.\nonumber\\
&&\left.-\mathcal C_q\mathcal S_q\mathcal C_{q+k}\mathcal S_{q+k}
[G_A(q,i\Omega) + G_B(-q,-i\Omega)] [G_A(q+k,i\Omega+i\omega)+
G_B(-q-k,-i\Omega-i\omega)]\right]\nonumber
\end{eqnarray}
with the notations $\mathcal C_q=\cos{\theta_q\over2}$ and
$\mathcal S_q=\sin{\theta_q\over2}$. Using the results
$G_A(k,i\omega)=G_B(k,i\omega)=G(k,i\omega)$ and
$G(-k,i\omega)=G(k,i\omega)=\frac{1}{i\omega-E_k}$, we get the
static susceptibility
\begin{eqnarray}
\chi_z&=&\lim_{k\to0}\chi_z(k,0)\nonumber\\
&=&\lim_{k\to0}\frac{2}{\beta V}\sum_{q,i\Omega}\left[-(\mathcal
C_q\mathcal C_{q+k}+\mathcal S_q\mathcal S_{q+k})^2
G(q,i\Omega)G(q+k,i\Omega) +(\mathcal C_q\mathcal S_{q+k}-\mathcal
S_q\mathcal C_{q+k})^2
G(q,i\Omega)G(q+k,-i\Omega)\right]\nonumber\\
&=&\lim_{k\to0}\frac{2}{V}\sum_{q}\left[-(\mathcal C_q\mathcal
C_{q+k}+\mathcal S_q\mathcal S_{q+k})^2
\frac{n_f(\varepsilon_{q+k})-n_F(\varepsilon_q)}{\varepsilon_{q+k}
-\varepsilon_q}\right]\nonumber\\
&=&-2\int d^2q\frac{\partial n_F(\varepsilon_q)}{\partial
\varepsilon_q}
\end{eqnarray}
which is the Pauli susceptibility for free fermions. At low
temperature, the excitation spectrum can be approximated by Dirac
cones, so we have
\begin{eqnarray}
\chi_z&\doteq&-2\int_0^\infty 2\pi q\frac{\partial
n_F(\varepsilon_q)}{v_F\partial q}dq\nonumber\\
&=&\frac{4\pi}{v_F}\int_0^\infty
n_F(\varepsilon_q)dq\propto\frac{T}{v_F^2}
\end{eqnarray}
\end{widetext}



\bigskip
\noindent

\begin{thebibliography}{99}

\bibitem{spinwave} J. Van Kranendonk and J. H. Van Vleck, Rev. Mod. Phys. 30, 1
(1958).

\bibitem{1/S}Jun-ichi Igarashi, Phys. Rev. B 46, 10763 (1992).

\bibitem{Haldane} F. D. M. Haldane, Physics Letters, 93 A, 464
(1983); Phys. Rev. Lett. 50, 1153 (1983).

\bibitem{AndersonSL} P. W. Anderson, Mater. Res. Bull. 8, 153 (1973);
Science 235, 1196 (1987).


\bibitem{Leescience} Patrick. A. Lee, Science, 321, 1306 (2008).

\bibitem{MF} D. P. Arovas and Assa Auerbach, Phys. rev. B 38, 316
(1988).

\bibitem{Ma} Sanjoy Sarker, C. Jayaprakash, H. R. Krishnamurthy and Michael Ma,
Phys. Rev. B 40, 5028 (1989).

\bibitem{AZHA-1/2} I. Affleck, Z. Zou, T. Hsu and P. W. Anderson,
Phys. Rev. B 38, 745 (1988).


\bibitem{Affleck&Marston} I. Affleck, Phys. Rev. Lett. 54, 966 (1985);
I. Affleck and J. B. Marston, Phys. Rev. B 37, 3774 (1988).

\bibitem{Read&Sachdev}N. Read, Subir Sachdev, Nucl. Phys. B 316, 609
(1989).

\bibitem{JPA}Guang-Ming Zhang and Xiao-guang Wang, J. Phys. A 39, 8515
(2006).

\bibitem{PRB}Hong-Hao Tu, Guang-Ming Zhang, Tao Xiang, Zheng-Xin Liu and Tai-Kai Ng,  Phys. Rev. B  80, 014401
(2009).

\bibitem{S=1/2}Actually, a similar expression like Eq.~(\ref{H3/2}) also holds for $S=1/2$,
$\hat{\mathbf{S}}_i\cdot\hat{\mathbf{S}}_j
=-\frac{J}{4}:[-\mathrm{Tr}(\psi_i^\dagger\mathbf{I}\psi_j\cdot\psi_j
^\dagger\mathbf{I}\psi_i)+ S^2\mathrm{Tr}(\psi_i^\dagger\psi_j
\psi_j^\dagger\psi_i)]:$. Notice that the first term in the square
bracket has different sign comparing to the expression for $S=1$
and 3/2. The two terms in the square bracket are also identical.

\bibitem{cheng&Ng}Chi-Ho Cheng and Tai-Kai Ng, Europhys. Lett., 52 (1), p. 87
(2000).

\bibitem{g4} Patrick A. Lee, Naoto Nagaosa, Tai-Kai Ng and Xiao-Gang Wen, Phys. Rev. B, 57, 6003
(1998).

\bibitem{LeeLee} Sung-Sik Lee and Patrick A. Lee, Phys. Rev. Lett 95, 036403
(2005).

\bibitem{MSpinWave} We note that another method to deal with disordered
spin states is modified spin-wave theory where zero magnetization is
enforced by a magnon number constraint. Minoru Takahashi [Prog.
Theor. Phys. Suppl. 87, 233 (1986); Phys. Rev. B 40, 2494 (1989)]
applied this method to study 1D ferromagnets and 2D antiferromagnets
on square lattice. The results agree well with that of Bethe Ansatz
and Schwinger boson mean field theory, respectively. This is because
that the elementary excitations in both situations are spin-1
magnons which can be viewed as bound states of two spin-1/2 spinons.
However, for 1D half-odd-integer antiferromagnetic spin chains,
elementary exciations are spin-1/2 spinons. So far we do not know
how this modified spin wave-theory can be generalized to deal with
this situation.

\bibitem{science}Satoru Nakatsuji, Yusuke Nambu, Hiroshi Tonomura, Osamu Sakai
Seth Jonas, Collin Broholm, Hirokazu Tsunetsugu, Yiming Qiu and
Yoshiteru Maeno, science, 309 ,1697 (2005).

\bibitem{PRBexp}Subhro Bhattacharjee, Vijay B. Shenoy and T. Senthil,  Phys. Rev. B, 74, 092406
(2006).

\bibitem{AFQ} Hirokazu TSUNETSUGU and M. ARIKAWA, J. Phys. Soc. Jpn. 75, 083701
(2006).

\bibitem{SU(3)} Peng Li, Guang-Ming Zhang and Shun-Qing Shen, Phys. Rev. B 75, 104420 (2007).

\bibitem{Qphase} A. L\"{a}uchli, Fr\'ed\'eric Mila and Karlo Penc,
Phys. Rev. Lett. 97, 087205 (2006).

\bibitem{J1J3K}E. M. Stoudenmire, Simon Trebst and Leon Balents, Phys. Rev. B 79,
214436 (2009).

\bibitem{Vorex}Hikaru KAWAMURA and A. YAMAMOTO, J. Phys. Soc. Jpn.
76, 073704 (2007); Chyh-Hong Chern, Phys. Rev. B 78, 020403 (2008).

\bibitem{Liu} Zheng-Xin Liu, Yi Zhou and Tai-Kai Ng, Phys. Rev. B 81, 224417 (2010).

\bibitem{Ng} Tai-Kai Ng, Phys. Rev. B, 50 555(1994).

\bibitem{Read} N. Read and D. Green, Phys. Rev. B, 61, 10267 (2000).

\bibitem{Tinkham} M.Tinkham, $Introduction\ to\ Superconductivity$,
New York (1975).

\bibitem{dimer}Ian Affleck, Nucl. Phys. B 265,
409 (1986); S. K. Yip, Phys. Rev. Lett. 90, 250402(2003); J. J.
Garc\'{i}a-Ripoll, M. A. Martin-Delgado, and J. I. Cirac, Phys. Rev.
Lett. 93, 250405 (2004); Rajiv R. P. Singh, Zheng Weihong, C. J.
Hamer, and J. Oitmaa, Phys. Rev. B 60, 7278 (1999).

\bibitem{Z2} T. Senthil and Matthew P. A. Fisher, Phys. Rev. B 62, 7850 (2000).

\bibitem{Hermele} M. Hermele, T. Senthil, Matthew P. A. Fisher, Patrick A. Lee,
Naoto Nagaosa, and Xiao-Gang Wen, Phys. Rev. B 70, 214437 (2004).


\end{thebibliography}
\end{document}